**Classification:** PHYSICAL SCIENCES-Engineering

# Origami Lattices and Folding-induced Lattice Transformations

**Hongbin Fang[a,b,1], Suyi Li[c], Manoj Thota[a], and K.W. Wang[a]**

**Affiliations:**

[a] Department of Mechanical Engineering, University of Michigan, Ann Arbor, MI 48109, USA

[b] Institute of AI and Robotics, Fudan University, Shanghai 200433, China

[c] Department of Mechanical Engineering, Clemson University, Clemson, SC 29634, USA

**[1]To whom correspondence should be addressed.**

| | |
|---|---|
| **Current address:** | Institute of AI and Robotics, Fudan University, 220 Handan Rd., Yangpu District, Shanghai 200433, China. |
| **Email:** | hongbinf@umich.edu and fanghongbin@fudan.edu.cn. |
| **Tel.:** | +86-13601887811 |

**ORCID identifiers:**

| | |
|---|---|
| **Hongbin Fang** | 0000-0001-6691-0531 |
| **Suyi Li** | 0000-0002-0355-1655 |
| **Manoj Thota** | 0000-0003-1894-6710 |
| **Kon-Well Wang** | 0000-0002-9163-2614 |

**Short title:** Origami Lattices and Lattice Transformations

**The authors declare no conflict of interest.**

**Abstract:** Lattices and their underlying symmetries play a central role in determining the physical properties and applications of many natural and engineered materials. By bridging the lattice geometry and rigid-folding kinematics, this study elucidates that origami offers a comprehensive solution to a long-standing challenge regarding the lattice-based materials: how to systematically construct a lattice and transform it among different symmetric configurations in a predictable, scalable, and reversible way? Based on a group of origamis consisting of generic degree-4 vertices, we successfully construct all types of 2D and 3D Bravais lattices, and demonstrate that they can undergo all diffusionless phase transformations via rigid-folding (i.e., dilation, extension, contraction, shear, and shuffle). Such folding-induced lattice transformations can trigger fundamental lattice-symmetry switches, which can either maintain or reconstruct the nearest neighborhood relationships according to a continuous symmetry measure. This study can foster the next generation of transformable lattice structures and materials with on-demand property tuning capabilities.

Lattices and their underlying symmetries play a central role in determining physical properties – such as band structure, compressibility, and elastic modulus – of many natural (1, 2) and engineered materials (3, 4). Such lattice-property relationship is particularly evident in the metamaterials (5–7). Therefore, purposefully designing and adjusting lattice topology can significantly expand the achievable material property range (8–10). However, constructing lattice structures from the ground-up is quite challenging, and once the material is synthesized, its constituent lattice typically cannot be modified. There is a lack of an integrated and scalable approach for constructing and reconfiguring lattice structures on-demand, let alone transforming their symmetry properties. Here we demonstrate that origami folding offers a solution to fill this gap.

Origami has become a popular subject among mathematicians, educators, physicists, and engineers owing to the seemingly infinite possibilities of transforming two-dimensional (2D) sheets into three-dimensional (3D) shapes via folding (11–15). Historically, such folding-induced shape transformations have been examined based on the spatial positions and orientations of its *facet surfaces* and *crease lines* (16, 17). For example, many origami-based mechanical metamaterials are analyzed by considering the folding as coordinated facet rotations with respect to the hinge-like creases – essentially a linkage mechanism (18, 19). Here, we examine the origami folding through a different lens by asking: How folding can spatially arrange and re-arrange the *characteristic entities* in the origami? These characteristic entities can be the vertices where crease lines intersect, or the center points of crease lines and facets. By treating these entities as the elements of a lattice (aka. lattice points), we uncover that origami offers a remarkably comprehensive framework to construct all 2D and 3D Bravais lattices, induce all types of diffusionless lattice transformations (aka. dilation, extension, contraction, shear, and shuffle), switch the lattice structure among different symmetric configurations, and even reconstruct the nearest neighbor relationships based on a continuous symmetry measure. Therefore, this result can fundamentally advance our capability to engineer high-performance material systems by incorporating the rich design and kinematics of origami folding.

# Results

## Origami lattice constructions

Origami structures are versatile scaffolds for constructing 2D and 3D lattices. Here we selectively assign lattice points either at the facet centers (Fig. 1, A and D) or vertices (Fig. 1, B and E). These lattice points can be occupied by inclusions of the same type (Fig. 1, A, C, and D) or different types (Fig. 1, B and E). The inclusions can be selected according to targeted applications, such as conducting element (20) and sonic barrier rods (21, 22). Assembly of multiple origami sheets can further enrich the 3D design space (Fig. 1, C, D, and E). By establishing correlations between the origami geometry and the lattice configuration, we discover that all 5 types of 2D and 14 types of 3D Bravais lattices – a well-established description of the lattice configuration according to symmetry – can be constructed on a group of simple and rigid-foldable origamis consisting of generic degree-4 vertices (23, 24) (*SI Appendix*, Fig. S1-S4, and Table S1, S2). The mapping between origami geometry and lattice configuration is not one-to-one in that multiple origami designs and lattice point arrangements can satisfy the reflection and rotational symmetry requirements of one specific Bravais lattice.

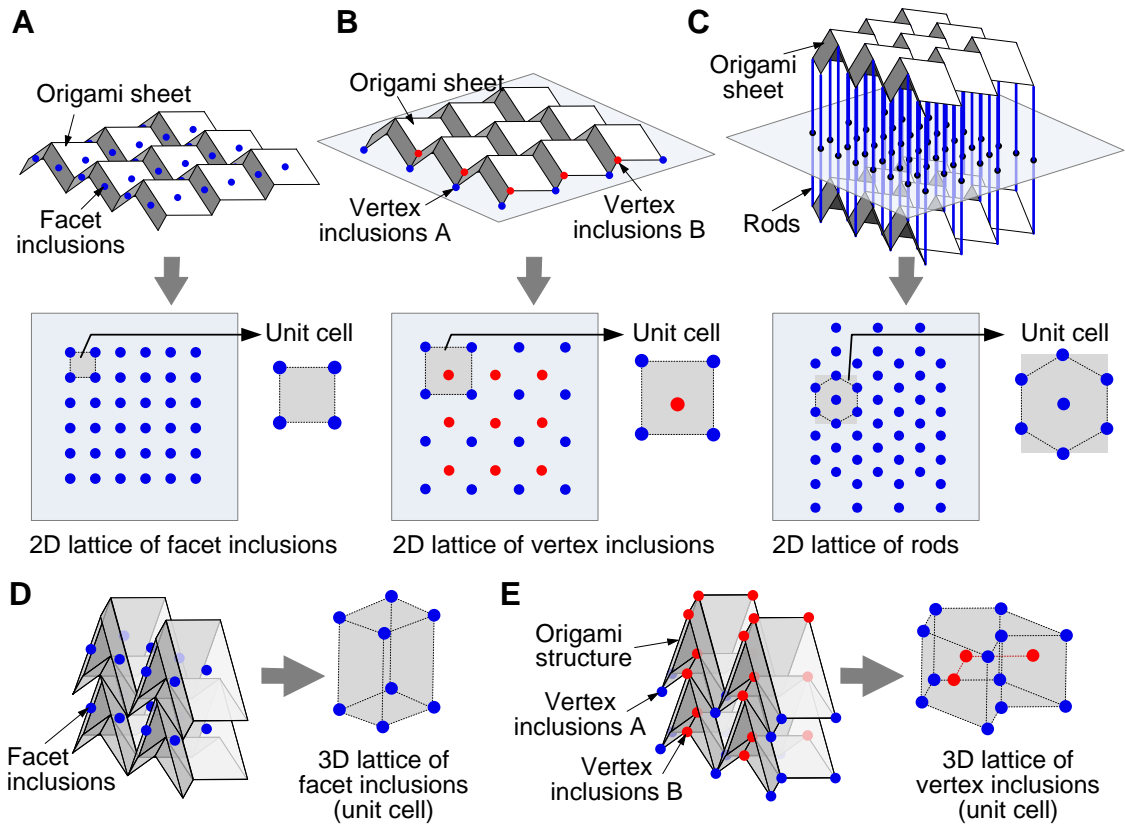

**Fig. 1. Construction of origami lattices of inclusions.** As examples, 2D lattices of (**A**) facet inclusions (square lattice), (**B**) vertex inclusions (non-Bravais lattice), and (**C**) rods (hexagonal lattice) are generated based on Miura-ori sheets; 3D lattices of (**D**) facets inclusions (primitive orthorhombic lattice) and (**E**) vertex inclusions (non-Bravais lattice) are generated based on stacked Miura-ori structures. Unit cells of these lattices are highlighted.

**Folding-induced diffusionless phase transformation and discrete symmetry switches**

Besides being a versatile platform for lattice construction, origami also provides effective mechanisms to transform these lattices via folding. By translating the folding-induced facet and vertex re-arrangement into the corresponding lattice point movements (Fig. 2), we discover that folding can impart all kinds of diffusionless phase transformations via attaining both lattice-distortive strains and shuffles. The former further includes dilation, contraction/extension, and shear (25) (*SI Appendix*, Table S3, S4). For 2D lattices of vertex inclusions, we show that origami sheets with negative in-plane Poisson's ratio (e.g., Miura-ori (19)) can induce 2D dilations (Fig. 2A); those with positive in-plane Poisson's ratio (e.g., eggbox pattern (26)) can bring about 2D contractions/extensions (Fig. 2B); and those with in-plane shear deformations (e.g., single-collinear (SC) origami (24)) can trigger 2D shear components (Fig. 2C). 2D shuffle, manifested as a small movement of a lattice point (red) within the unit cell, can also be achieved in a Miura-ori-based lattice (Fig. 2D). For 3D lattices of vertex inclusions, 3D dilation is achievable by exploiting the tri-directional auxeticity of a stacked general flat-foldable (GFF) structure (24) (Fig. 2E); 3D contraction/extension is a result of the combination of positive and negative Poisson's ratios in a stacked Miura-ori structure (19) (Fig. 2F); 3D shear is achievable due to the out-of-plane shear deformation in a GFF sheet (Fig. 2G) (24); and 3D shuffle is evident when the lattice point locating at the center vertex of the Miura-ori cell translates within the 3D unit lattice cell during folding (Fig. 2H). It is worth noting that the shear or shuffle components do not occur alone, and they are always accompanied by dilation or contraction/expansion. These rich connections between origami kinematics and lattice-point movements can enable us to program a broad range of lattice transformations – according to the Cohen's classification for example (27) – by purposefully designing the crease patterns and their folding motions.

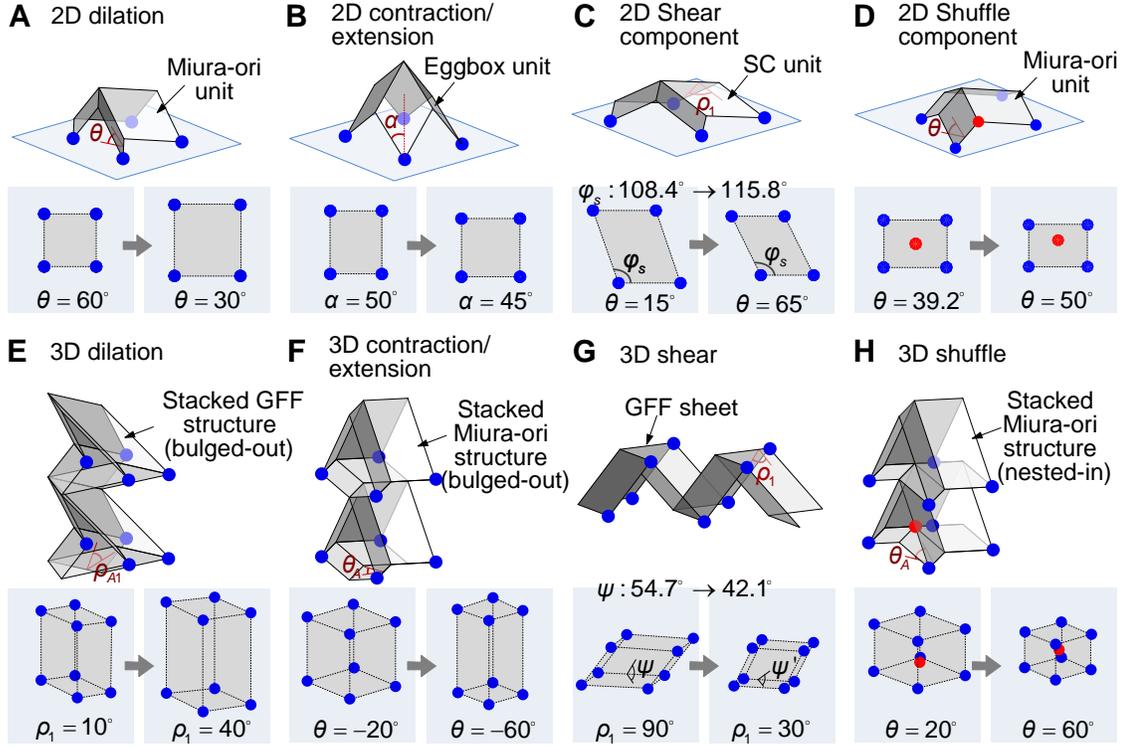

**Fig. 2. Diffusionless phase transformations in origami lattices.** Folding of origami structures and the corresponding lattice transformations for (**A**) 2D dilation, (**B**) 2D contraction/extension, (**C**) 2D shear component, (**D**) 2D shuffle component, (**E**) 3D dilation, (**F**) 3D contraction/extension, (**G**) 3D shear component, and (**H**) 3D shuffle component. Folding is represented as the change of certain geometric angle. Detailed geometries and kinematics of these origami structures can be found in *SI Appendix* ,Table S3, and Fig. S5, S6.

More importantly, these folding-induced diffusionless transformations can fundamentally switch the lattices' underlying symmetry groups. For instance, the 2D contraction/extension could add an additional reflection symmetry to the rectangular lattice based on Miura-ori and switch it into a square lattice (i.e., the symmetry group switches from **D₂** to **D₄**, Fig. 2B). The 2D shuffle, on the other hand, could break down a reflection symmetry of a centered rectangular lattice and evolve it into a non-Bravais lattice (**D₂** to **D₁**, Fig. 2D). Remarkably, with a carefully designed crease pattern, origami folding can switch the corresponding lattice among more than two Bravais types, which could appear either alternatively or repeatedly. Figure 3A shows that by prescribing the crease design parameters ($a$, $c$, and $\gamma$), the Miura-ori based 2D lattices of rods could reach four types of 2D Bravais lattice: rectangular (R), center-rectangular (CR), square (S), and hexagonal (H) lattices [21]. For example, if $c/a=1$ and $\gamma=60°$, the Miura-ori-based lattice of rods will undergo a series of discrete switches: from being a hexagonal lattice ($\theta = 0^{\circ}$) to a square lattice ($\theta = 54.7^{\circ}$), to a hexagonal lattice again ($\theta = 70.5^{\circ}$), and finally to a rectangular lattice ($\theta \rightarrow 90^{\circ}$) (Fig. 3B). Note that between these switches, the origami lattices are non-Bravais, which are not denoted in Fig. 3B. Detailed parametric analysis and other examples of discrete symmetry switches can be found in *SI Appendix* and Fig. S7.

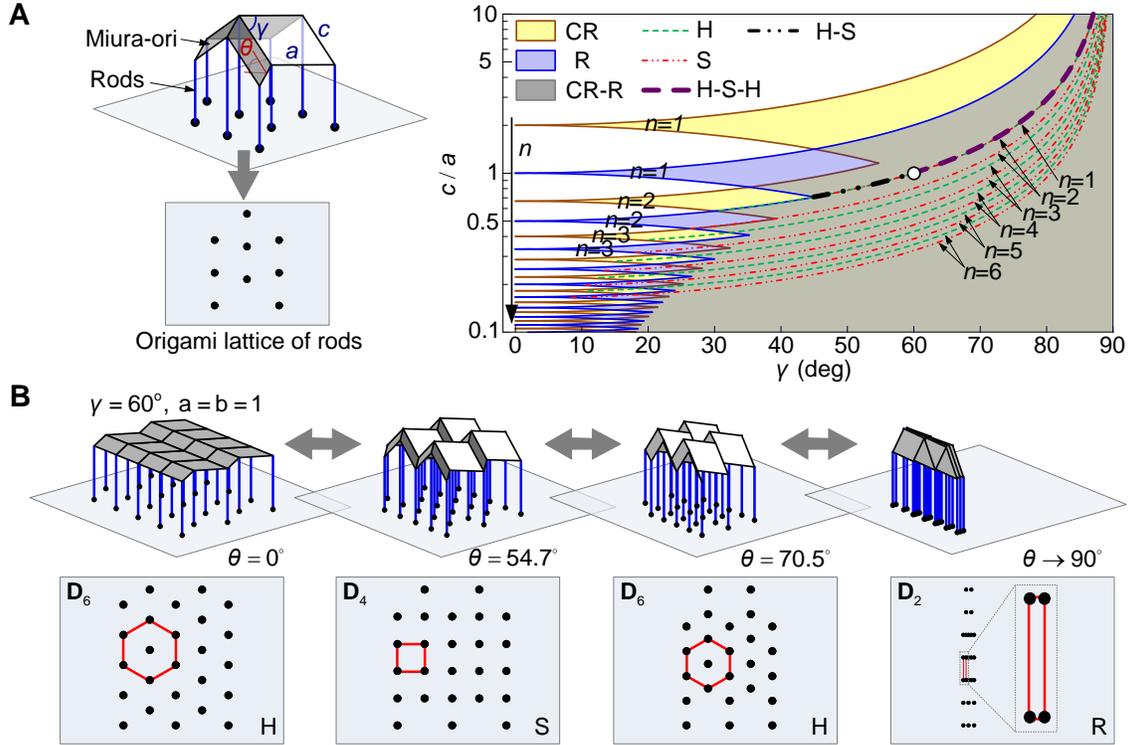

**Fig. 3. Discrete symmetry switches of origami lattices.** (A) The 2D lattice of rods based on Miura-ori sheet could switch among different 2D Bravais lattice types. The contour plot shows the achievable Bravais lattices versus Miura-ori geometries, where $n$ is a positive integer denoting geometry relations, see details in *SI Appendix* and Fig. S7. Based on $c/a = 1, \gamma = 60^{\circ}$ (shown as the circle in the contour plot), (B) demonstrates the discrete switches of the lattice from a hexagonal, to a square, to another hexagonal, and finally to a rectangular type. Correspondingly, the symmetry switches from $\mathbf{D_6}$, to $\mathbf{D_4}$, to $\mathbf{D_6}$, and finally to $\mathbf{D_2}$.

**Folding-induced continuous symmetry evolutions**

So far, we have been focusing on discrete folding configurations where the corresponding lattices are *strictly* symmetric. However, folding is a continuous process; when the origami is folded slightly away from these strictly symmetric configurations, the corresponding lattice would no longer possesses certain symmetry in a strict sense but still very close. Therefore, it is necessary to derive a quantitative method to analyze the folding-induced and continuous changes in the lattice symmetry. Here we adopt the concept of *continuous symmetry measure* (CSM), which was first proposed in the field of structural chemistry (28, 29). Given a shape and a symmetry group **G** (such as $\mathbf{D_2}$ and $\mathbf{D_4}$), the continuous symmetry measure, $S(\mathbf{G})$, quantifies the minimal displacement that the points of the object have to undergo in order to be transformed into a shape with **G**-symmetry. Hence, an object features a zero $S(\mathbf{G})$ when it is strictly **G**-symmetric; if this object has to undergo larger displacements to acquire **G**-symmetry, its $S(\mathbf{G})$ value increases accordingly. In other words, rather than a "black and white" discrete approach to describe symmetry (that is, either symmetric or asymmetric), the CSM offers a more continuous "grey" scale to characterize the strength of a particular type of symmetry group. Detailed mathematical description of this concept and its application in quantifying origami lattice can be found in the *SI Appendix* and (28, 30). By adopting this measure, we are able to describe how much the origami lattice possesses certain symmetry and understand its nearest symmetric correspondence at certain folding configuration.

As an example, we revisit the case in Fig. 3B by evaluating the CSM of the origami lattice with respect to symmetry groups $\mathbf{D_6}$, $\mathbf{D_4}$, and $\mathbf{D_2}$, which correspond to the hexagonal, square, and rectangular (or center rectangular) lattices, respectively. We select a hexagon unit cell (i) and a parallelogram unit cell (ii) of the lattice and calculate their CSM values with respect to the different symmetry groups throughout the folding range from $\theta = 0°$ (fully flat) to $\theta = 90°$ (fully folded) (Fig. 4A, see detailed calculation procedures in the *SI Appendix* and another example in Fig. S8). At $\theta = 0°$, $S_i(\mathbf{D_6}) = 0$ for cell (i), meaning that the lattice is of hexagonal type and strictly exhibits $\mathbf{D_6}$ symmetry. As $\theta$ increases, $S_i(\mathbf{D_6})$ gradually increases, indicating a loss of $\mathbf{D_6}$ symmetry; however, $S_{ii}(\mathbf{D_4})$ decreases for cell (ii), which is a signal of strengthening $\mathbf{D_4}$ symmetry. At $\theta = 54.7°$, $S_i(\mathbf{D_6})$ reaches a local maxima whereas $S_{ii}(\mathbf{D_4}) = 0$, this suggests that the lattice is far away from $\mathbf{D_6}$ symmetry but fully acquires the $\mathbf{D_4}$ symmetry (aka. square lattice). As $\theta$ further increases, $S_i(\mathbf{D_6})$ starts to decrease but $S_{ii}(\mathbf{D_4})$ increases. At $\theta = 70.5°$, $S_i(\mathbf{D_6})$ returns to zero while $S_{ii}(\mathbf{D_4})$ reaches a local maxima, implying that the lattice fully regains $\mathbf{D_6}$ and loses $\mathbf{D_4}$ symmetry. At the final stage of folding, $S_{ii}(\mathbf{D_2})$ converges to zero while $S_{ii}(\mathbf{D_4})$ climbs quickly, indicating an emergence of a rectangular lattice with $\mathbf{D_2}$ symmetry when $\theta$ approaches $90°$.

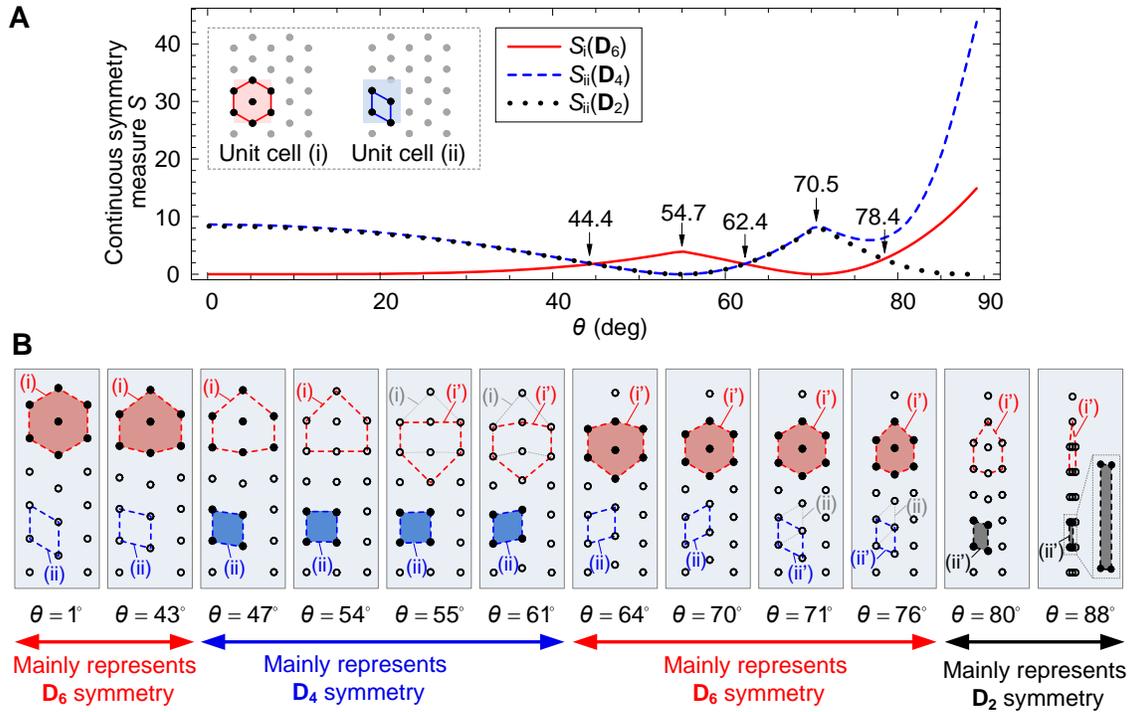

**Fig. 4. Continuous symmetry measures and reconstruction of the nearest-neighbor relationship of the origami lattices.** Based on the same design as Fig. 3B, (A) displays the CSM values of the two unit cells (i) and (ii) with respect to symmetry groups $\mathbf{D_6}$, $\mathbf{D_4}$, and $\mathbf{D_2}$, where the arrows denote the reconstructions of the nearest-neighbor relationship. For clear illustration, lattices at specific folding configurations are provided in (B). The unit cell (i) deforms via weak transformations when $\theta < 54.7^\circ$; as the folding reaches and passes $\theta = 54.7^\circ$, a reconstructive transformation occurs and a new unit cell (i') with reconstructed nearest lattice-point neighbors is generated. Similar reconstructive transformation happens on unit cell (ii) when the folding reaches and passes $\theta = 70.5^\circ$, with a new unit cell (ii') being generated. If examining the overall lattice, we observe another type of reconstructive transformation. When reaching and passing $\theta = 44.4^\circ$, the nearest neighbors changes from those lattice points constituting the unit (i) to those constituting the unit cell (ii); the lattice mainly represents $\mathbf{D_6}$ and $\mathbf{D_4}$ symmetry before and after $\theta = 44.4^\circ$, respectively. Similar reconstructions also occur at $\theta = 62.4^\circ$ and $78.4^\circ$. At each folding configuration, the lattice with the nearest neighbors is denoted by shaded polygon with solid lattice points.

## Nearest-neighbor relationship of the origami lattices

To understand the physics underpinning the folding-induced discrete symmetry switches and CSM evolutions, we carefully examine the lattice point movements during these transformations. Around those configurations with strict symmetry, even with a relatively small range of folding, the lattice point movements can be sufficient to switch the symmetry group but without breaking the nearest-neighbor relationships. For example, the contraction/extension shown in Fig. 2B can switch the symmetry group from $\mathbf{D}_2$ to $\mathbf{D}_4$ by slightly moving the 4 lattice points in the unit cell such that the corresponding rectangle changes to a square, but the unit cell always consists of the same 4 lattice points during this process. Such transformation mechanism that maintains the nearest-neighbor relationship is defined as "*weak*", which is manifested as smooth variations of the CSM value in Fig. 4A. However, near some critical folding configurations, the lattice points within the unit cell moves in a way that they are no longer the nearest neighbors, so that a new unit cell of the same type is generated by incorporating a different set of lattice points. Such mechanism involving breaking and reconstructing of nearest-neighbor relationships is defined as "*reconstructive*"[*]. For example in Fig. 4B, the hexagonal shape formed by the lattice points of unit cell (i) first undergoes a weak transformation, from $\theta = 0°$ to near $54.7°$, as it becomes a house-like shape. When $\theta$ reaches and passes $54.7°$, one lattice points of the unit cell is replaced by a new one such that a new unit cell with the nearest lattice-point neighbors is reconstructed. In the CSM plot, such reconstructive transformation is manifested as the non-smooth maxima corresponding to this particular unit cell (e.g., at $\theta = 54.7°$ for unit cell (i) and $\theta = 70.5°$ for unit cell (ii)).

The reconstructive mechanisms discussed so far are defined on a particular type of unit cell, and we can extend its definition by considering different types of unit cells in the origami lattice simultaneously. At any folding configurations, the unit cell with relatively lower value of the CSM better represents the prominent symmetry of the overall lattice. Therefore, among different types of unit cell, when there is a change in terms of the smallest CSM, the nearest-neighbor relationship of the lattice can be also considered to experience a reconstruction. The difference here is that such a reconstruction occurs between unit cells of different types. In the example shown in Fig. 4, $S_i(\mathbf{D}_6)$ is lower than $S_{ii}(\mathbf{D}_4)$ before $\theta$ reaches $44.4°$, indicating that the lattice points in the unit cell (i) constitutes the nearest-neighbor relationship of the overall lattice, and the overall lattice mainly represents $\mathbf{D}_6$ symmetry; however, $S_i(\mathbf{D}_6)$ outstrips $S_{ii}(\mathbf{D}_4)$ as $\theta$ passes $44.4°$, suggesting that the lattice points in the unit cell (ii) become the nearest neighbors, and the overall lattice mainly exhibits $\mathbf{D}_4$ symmetry. Based on this generalized definition of the reconstructive mechanism, it is evident from Fig. 4 that folding can trigger multiple and successive reconstructions of the nearest-neighbor relationships (e.g., at $\theta = 44.4°$,

---

[*] It is worth noting that here 'reconstructive' does not mean 'diffusional' but rather denotes the broken and regeneration of the nearest-neighbor relationship; all transformations discussed in this paper are diffusionless.

$\theta = 62.4°$, and $\theta = 78.4°$).

3D Bravais lattices constructed on the generic 4-vertex origami can also exhibit symmetry switches and evolutions by folding. We find that lattices of the same centering type (primitive, body-centered, base-centered, and face-centered lattices) possess the potential to evolve among different crystal families, which need further exploration. It is also worth highlighting that unlike certain martensitic transformations that are irreversible (1, 2), the above origami folding-induced transformations are always reversible even after multiple reconstructions of the nearest-neighbor relationship.

**Conclusions**

This study uncovers and elucidates the comprehensiveness of the capability to construct and reconfigure lattice structures by origami. All 5 types of 2D and 14 types of 3D Bravais lattices can be constructed by exploiting the design space of origami and carefully selecting its characteristic points for lattice-point arrangement. All 2D and 3D diffusionless lattice transformations – including dilation, extension/contraction, shear, and shuffle – can be achieved in such origami lattices via rigid folding. More importantly, we discover that folding could induce continuous and reversible conversions of the lattice symmetry, and these symmetry conversions can either maintain or reconstruct the nearest-neighbor relationship in a predicable manner.

Since the lattice structure and its symmetry properties directly govern many physical properties, the origami lattice construction and transformation offers us great freedom to architect programmable and adaptive metamaterials. Here the lattice points can be occupied by any components used for metamaterials, including geometric entities (9, 31), mechanical units (32), acoustic modules (33), electromagnetic devices (7, 20), and photonic elements (7, 34), etc. Attaching these components to the origami scaffold gives us metamaterials and metastructures (from nanometer-scale DNA origami (35) to meter-scale origami space structure (13)) whose mechanical, thermal, acoustic, optical, and electromagnetic properties can be effectively tailored by folding on demand. The universality, versatility, and continuity of origami folding would significantly advance the state of the art by achieving the long-desired online controllability of the basic architectures.

**Materials and Methods**

The continuous symmetry measures (CSM) are evaluated based on the *Folding/Unfolding* method. Brief introductions to the CSM of origami lattices and the Folding/Unfolding method are provided in the *SI Appendix*, Figure S8, and S9. Interested readers can find more detailed mathematical principles and deviations of CSM in (28, 30).

# ACKNOWLEDGEMENT

The authors would like to thank Narayanan Kidambi for the helpful discussions. H. Fang, M. Thota, and K.W. Wang acknowledge the support from the National Science Foundation (Award #1634545) and the University of Michigan Collegiate Professorship. S. Li acknowledge the support from the National Science Foundation (Award # CMMI-1633952, 1751449 CAREER, 1760943) and Clemson University (startup funding and Dean's Faculty Fellow Award).

# SI Appendix for

## Origami Lattices and Folding-Induced Lattice Transformations

**Hongbin Fang[a,b,1], Suyi Li[c], Manoj Thota[a], and K.W. Wang[a]**

[a] Department of Mechanical Engineering, University of Michigan, Ann Arbor, MI 48109, USA
[b] Institute of AI and Robotics, Fudan University, Shanghai 200433, China
[c] Department of Mechanical Engineering, Clemson University, Clemson, SC 29634, USA

[1] To whom correspondence should be addressed:
Email: hongbinf@umich.edu and fanghongbin@fudan.edu.cn (H. Fang)

## Contents

## S1. Constructions of 2D Bravais lattices with degree-4 vertex origamis

There are 5 types of 2D Bravais lattices: oblique, rectangular, center rectangular, square, and hexagonal lattices. For simplicity, they are hereafter referred as 'O', 'R', 'CR', 'S', and 'H', respectively. In this section, we show how they can be constructed based on degree-4 vertex (4-vertex) origamis by providing the detailed correlations between origami geometry and lattice configuration. Note that there are multiple origami designs and characteristic entity assignments that can be used to create one specific type of 2D Bravais lattice. Thus, the construction given here serves as an example to elucidate the underlying geometric correlations.

To obtain an oblique lattice, a single-collinear (SC) origami pattern is used. Fig. S1A shows the geometry of an SC origami unit at both the flat and a folded state. An SC unit can be characterized by three crease lengths ($a$, $b$, and $c$) and two sector angles ($\gamma_1$ and $\gamma_2$) (left, Fig. S1A). Folding of the SC unit can be described by the folding angle $\theta$ $\left(\theta \in [0, 90^\circ]\right)$, which is defined as the dihedral angle between the facet '1-2-0-8' (or facet '0-6-7-8') and the reference $x$-$o$-$y$ plane. Vertices '1', '2', '3', '5', '6', and '7' are all coplanar in the $x$-$o$-$y$ plane. The outer dimensions of the SC unit are given by (1):

$$\xi = \arccos \frac{\cos \gamma_1}{\sqrt{1 - \sin^2 \gamma_1 \sin^2 \theta}}, \quad \eta = \arccos \frac{\cos \gamma_2}{\sqrt{1 - \sin^2 \gamma_1 \sin^2 \theta}}, \quad \varphi_S = \xi + \arccos \frac{a^2 + L^2 - b^2}{2aL},$$

$$L = \sqrt{a^2 + b^2 - 2ab\cos(\xi + \eta)}, \quad W = 2c\sqrt{1 - \sin^2 \gamma_1 \sin^2 \theta}, \quad H = c \sin \gamma_1 \sin \theta. \tag{S1}$$

where $H$, $W$, and $L$ are the height, width, and length of a folded SC unit, respectively (middle, Fig. S1A). By projecting the folded SC unit onto the reference $x$-$o$-$y$ plane, we obtain a polygon, whose shape are described by angles $\xi$, $\eta$, and $\varphi_S$ (middle and right, Fig. S1A).

The other four types of 2D Bravais lattices can be constructed by using Miura-ori patterns. Fig. S1B shows the geometry of a Miura-ori unit. A Miura-ori unit can be characterized by two crease lengths ($a$ and $c$) and one sector angle ($\gamma$) (left, Fig. S1B). Folding of the Miura-ori unit can also be described by the folding angle $\theta$ $\left(\theta \in [0, 90^\circ]\right)$ defined as the dihedral angle between one of its facets and the reference $x$-$o$-$y$ plane. Vertices '1', '2', '3', '5', '6', and '7' are coplanar in the $x$-$o$-$y$ plane. The outer dimensions of the Miura-ori unit are given by (2):

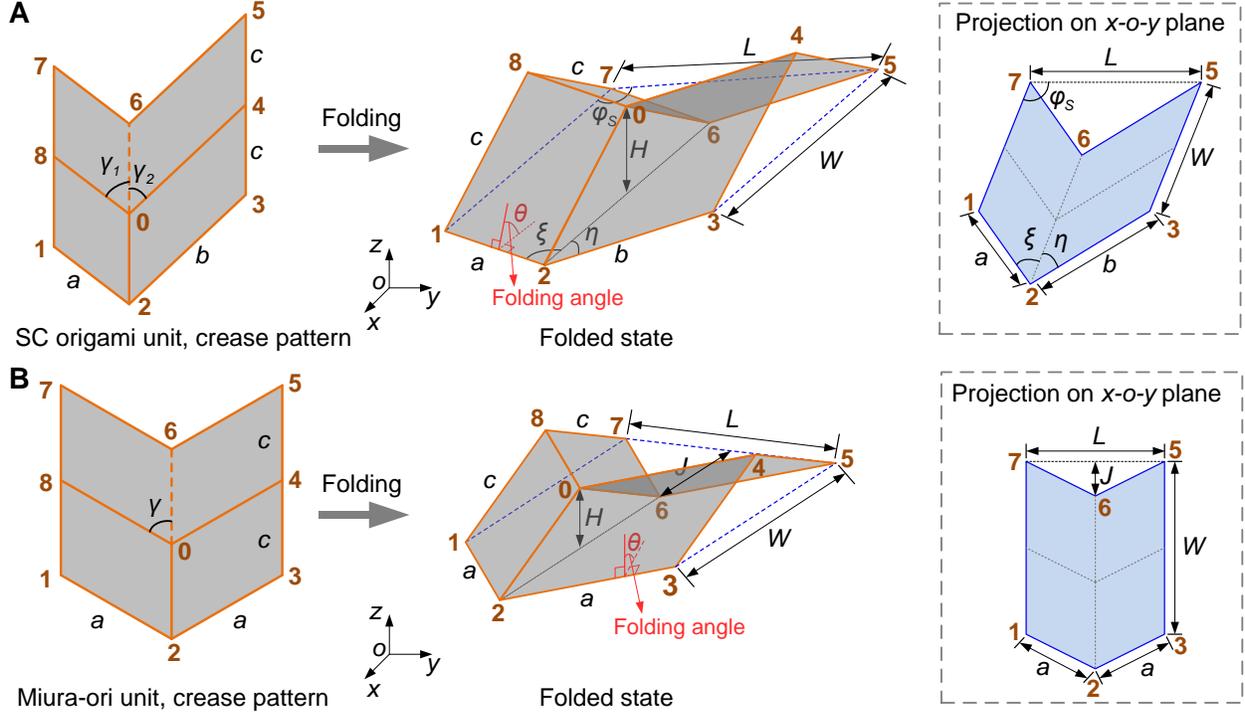

**Fig. S1. Geometries of 2D origami unit.** (**A**) An SC origami unit. (**B**) A Miura-ori unit. For each case, the crease pattern (left), the origami unit at a folded state (middle), and its projection onto the reference *x-o-y* plane (right) are given. In the crease patterns, 'mountain' and 'valley' creases are represented by solid and dashed lines, respectively.

$$H = c \sin \gamma \sin \theta, \quad W = 2c\sqrt{1 - \sin^2 \gamma \sin^2 \theta},$$
$$L = 2a \frac{\cos \theta \sin \gamma}{\sqrt{1 - \sin^2 \gamma \sin^2 \theta}}, \quad J = \frac{a}{\sqrt{1 + \tan^2 \gamma \cos^2 \theta}}, \tag{S2}$$

where $H$, $W$, $L$ are the height, width, and length of a folded Miura-ori unit, respectively (middle, Fig. S1B). Projecting the folded unit onto the *x-o-y* plane, a polygon is also obtained.

By treating the coplanar vertices '1', '2', '3', '5', '6', and '7' as lattice points and placing inclusions on them, we can construct 2D origami lattices of vertex inclusions. Fig. S2 shows the polygons that are projected from a folded origami unit and the corresponding 2D Bravais lattices. For each 2D Bravais lattice, the detailed correlations between the origami geometries and lattice geometries are given in Table S1.

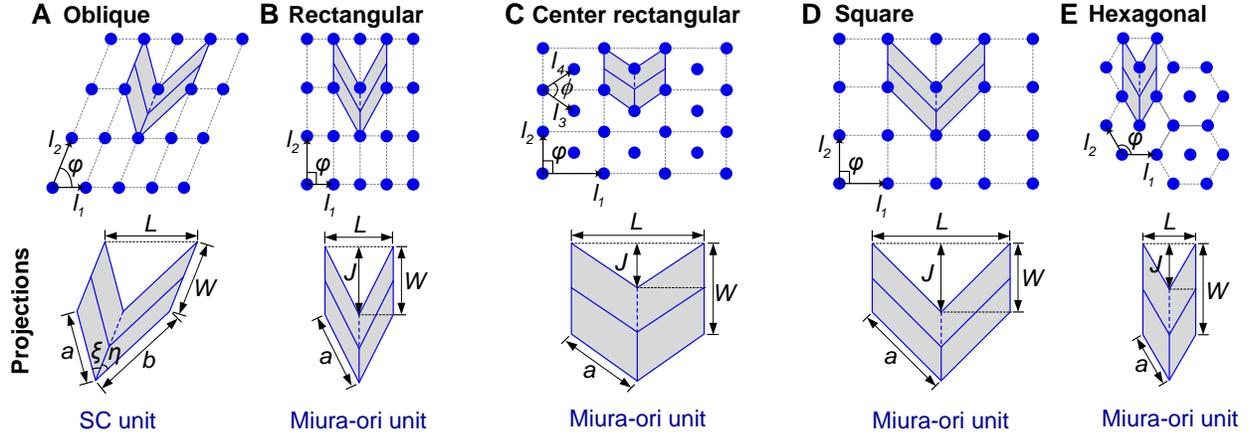

**Fig. S2. The 5 types of 2D Bravais lattices and the corresponding polygons projected from a folded origami unit**. (**A**) oblique lattice, (**B**) rectangular lattice, (**C**) center rectangular lattice, (**D**) square lattice, and (**E**) hexagonal lattice.

**Table S1.** Geometry correlations for constructing 2D Bravais lattices based on 4-vertex origamis

| 2D Bravais Lattices | Point Groups (symmetries) | Axial distances and axial angle | Origami patterns | Correlation between lattice geometry and origami geometry |
|---|---|---|---|---|
| **Oblique** (O) | $C_2$ | $l_1 \neq l_2$, $\varphi \neq 90°$ | SC | $a = \sqrt{l_1^2 + l_2^2 - 2l_1l_2\cos\varphi}$, $b = \sqrt{l_1^2 + l_2^2 + 2l_1l_2\cos\varphi}$, $W = l_2$, $L = 2l_1$, $\xi = \arcsin(l_1\sin\varphi / a)$, $\eta = \arcsin(l_1\sin\varphi / b)$ |
| **Rectangular** (R) | $D_2$ | $l_1 \neq l_2$, $\varphi = 90°$ | Miura-ori | $a = \sqrt{l_1^2 + l_2^2}$, $J = W = l_2$, $L = 2l_1$ |
| **Centered rectangular** (CR) | | $l_1 \neq l_2$, $\varphi = 90°$ ($l_3 \neq l_4$, $\phi \neq 90°$) | | $a = l_3$, $W = 2l_3\sin(\phi / 2)$, $J = W / 2$, $L = 2l_3\cos(\phi / 2)$ |
| **Square** (S) | $D_4$ | $l_1 = l_2$, $\varphi = 90°$ | | $a = \sqrt{2}l_1$, $J = W = l_1$, $L = 2l_1$ |
| **Hexagonal** (H) | $D_6$ | $l_1 = l_2$, $\varphi = 120°$ | | $a = l_1$, $W = \sqrt{3}l_1$, $J = W / 2$, $L = l_1$ |

## S2. Constructions of 3D Bravais lattices with degree-4 vertex origamis

3D Bravais lattices can be categorized into 7 systems. In each system, the lattice points in a unit cell can follow 4 different centering types. Specifically, the 7 systems are: triclinic, monoclinic, orthorhombic, tetragonal, rhombohedral, hexagonal, and cubic (3). The 4 centering types are: primitive, base-centered, body-centered, and face-centered. However, some combinations of lattice systems and centering types create the same lattice. After considering such redundancy, there are 14 unique types of 3D Bravais lattices.

We use three types of origami structures to construct these 3D Bravais lattices. They are: stacked SC structure in the bulged-out configuration, stacked Miura-ori structure in bulged-out configuration, and stacked double-collinear (DC) structure. In what follows, we formulate the geometries of these three stacked origami structures, and then detail the correlations between the origami geometry and 3D lattice configurations.

The stacked SC structure consists of two SC units (Fig. S3A). Each SC unit is characterized by crease lengths $a_x$, $b_x$, and $c_x$, and sector angles $\gamma_{x1}$ and $\gamma_{x2}$. Here the subscript '$x$' takes the values of '$A$' or '$B$' and represents the bottom unit A or the top unit B, respectively. Without loss of generality, we assume $\gamma_{x1} < \gamma_{x2}$. The following geometry constraints have to be reinforced to ensure the kinematic compatibility between the two SC units during the whole rigid-folding process (1):

$$a_A = a_B = a, \quad b_A = b_B = b, \quad \frac{\cos \gamma_{A1}}{\cos \gamma_{B1}} = \frac{\cos \gamma_{A2}}{\cos \gamma_{B2}} = \frac{c_B}{c_A}, \qquad (S3)$$

Folding of the stacked SC structure is still a single degree-of-freedom mechanism. It can be described by a folding angle $\theta_A$ (or $\theta_B$), which is defined as the dihedral angle between a facet of the bottom (or top) unit and the reference *x-o-y* plane (Fig. S3A). $\theta_A$ and $\theta_B$ are not independent to each other but satisfy the following relationship

$$\frac{\cos \theta_A}{\cos \theta_B} = \frac{\tan \gamma_{B1}}{\tan \gamma_{A1}}. \qquad (S4)$$

In the bulged-out configuration $-90° \le \theta_A \le 0$, and in the nested-in configuration $0 < \theta_A \le 90°$. The outer dimensions of the SC stacked structure when bulged-out are

$$H_A = c_A \sin \gamma_{A1} \sin |\theta_A|, \quad H_B = c_B \sin \gamma_{B1} \sin \theta_B, \quad H = H_A + H_B,$$

$$\xi = \arccos \frac{\cos \gamma_{A1}}{\sqrt{1 - \sin^2 \gamma_{A1} \sin^2 \theta_A}}, \quad \eta = \arccos \frac{\cos \gamma_{A2}}{\sqrt{1 - \sin^2 \gamma_{A1} \sin^2 \theta_A}},$$

$$W = 2c_A \sqrt{1 - \sin^2 \gamma_{A1} \sin^2 \theta_A}, \quad L = \sqrt{a_A{}^2 + b_A{}^2 - 2a_A b_A \cos(\xi + \eta)}, \tag{S5}$$

$$\varphi_S = \xi + \arccos \frac{a_A{}^2 + L^2 - b_A{}^2}{2a_A L}.$$

The stacked SC structure is used to construct triclinic and monoclinic lattices. For triclinic lattice, two additional angles $\varphi_1$ and $\varphi_2$ (Fig. S4A) are defined below to characterize the lattice

$$\varphi_1 = \arccos\left(-\frac{W \cos \varphi_S}{2a_A}\right), \quad \varphi_2 = \pi - \arcsin\left(\frac{\sqrt{c_A{}^2 - (W/2)^2}}{c_A}\right). \tag{S6}$$

The stacked Miura-ori structure consists of two Miura-ori units (Fig. S3B). Each Miura-ori unit is characterized by two crease lengths $a_x$ and $b_x$, and one sector angle $\gamma_x$, where the subscript '$x$' also takes the values of '$A$' or '$B$', denoting the bottom unit A or the top unit B, respectively. The following geometry constraints have to be reinforced to ensure the kinematic compatibility between the two units during rigid-folding (2):

$$a_A = a_B = a, \quad \frac{\cos \gamma_A}{\cos \gamma_B} = \frac{c_B}{c_A}, \tag{S7}$$

Similarly, folding of the stacked Miura-ori structure is a single degree-of-freedom mechanism. It can be described by a folding angle $\theta_A$ (or $\theta_B$), which is defined as the dihedral angle between a facet of the bottom (or top) unit and the reference $x$-$o$-$y$ plane (Fig. S3B). $\theta_A$ and $\theta_B$ are not independent to each other but satisfy the following relationship

$$\frac{\cos \theta_A}{\cos \theta_B} = \frac{\tan \gamma_B}{\tan \gamma_A}. \tag{S8}$$

In the bulged-out configuration $-90° \leq \theta_A \leq 0$, and in the nested-in configuration $0 < \theta_A \leq 90°$. The outer dimensions of the Miura-ori stacked structure when bulged-out are

$$H_A = c_A \sin \gamma_A \sin |\theta_A|, \quad H_B = c_B \sin \gamma_B \sin \theta_B, \quad H = H_A + H_B,$$

$$L = 2a_A \frac{\cos \theta_A \sin \gamma_A}{\sqrt{1 - \sin^2 \gamma_A \sin^2 \theta_A}}, \quad J = \frac{a_A}{\sqrt{1 + \tan^2 \gamma_A \cos^2 \theta_A}}, \quad W = 2c_A \sqrt{1 - \sin^2 \gamma_A \sin^2 \theta_A}. \tag{S9}$$

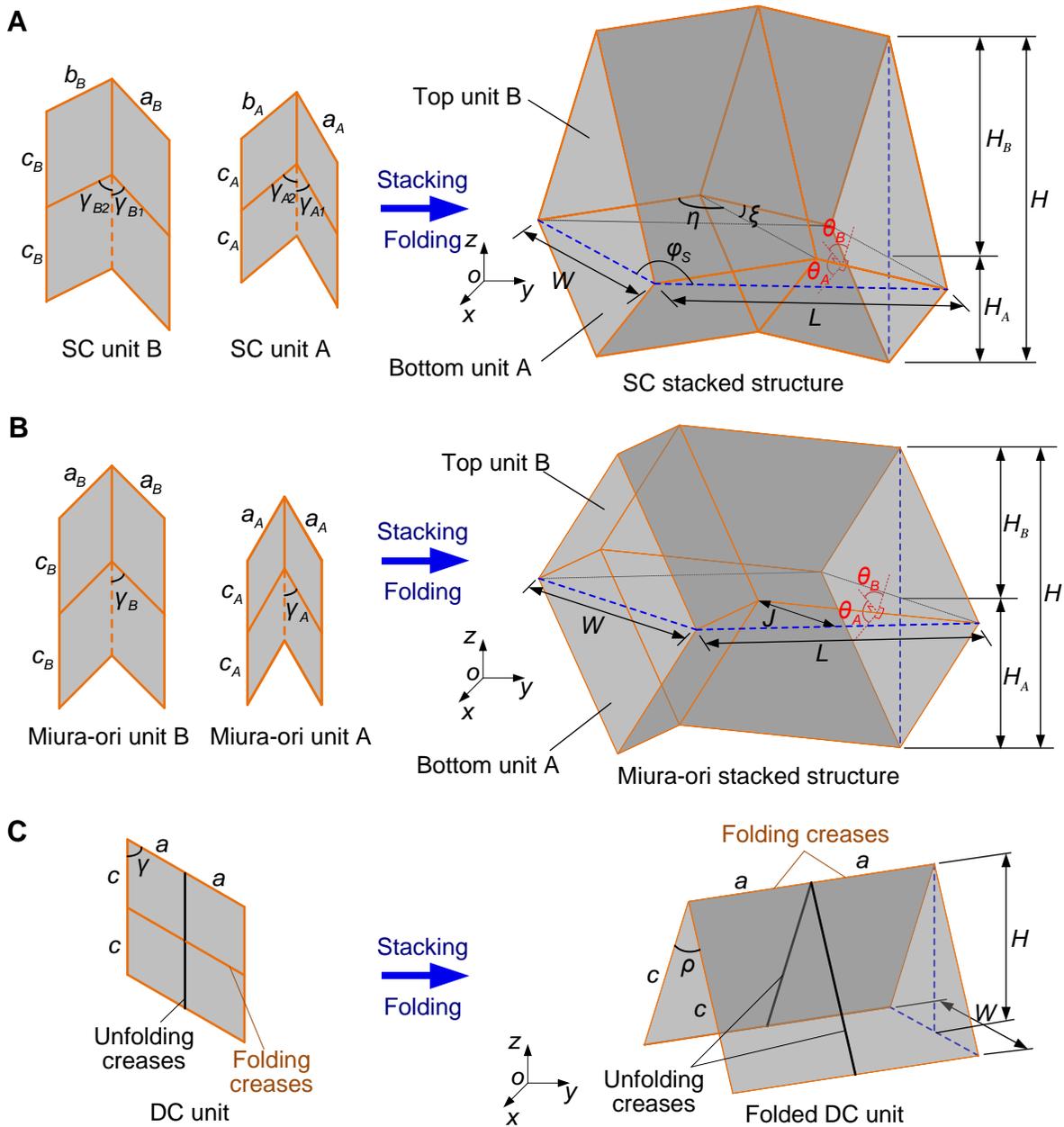

**Fig. S3. Geometries of 3D origami structures.** (**A**) A stacked SC structure (bulged-out), (**B**) a stacked Miura-ori structure (bulged-out), and (**C**) a folded DC unit. For each case, the constituent units' crease patterns (left) and the origami at a folded state (right) are given. For each crease pattern, the 'mountain' and 'valley' creases are denoted by solid and dashed lines, respectively. For the DC unit, the folded and unfolded creases are highlighted.

The stacked Miura-ori structure is used to construct orthorhombic, tetragonal, hexagonal, and cubic lattices.

The stacked DC structure is obtained by stacking two identical DC units (Fig. S3C). Each DC unit has two pairs of creases that are collinear, and it can be characterized by two crease lengths $a$, $c$ and one sector angle $\gamma$. Folding of a DC unit can be described by the dihedral folding angle $\rho$ between its facets and reference *x-o-y* plane. Hence, the outer dimensions of a folded DC unit are

$$H = c\cos(\rho/2), \; W = 2c\sin(\rho/2). \tag{S10}$$

The stacked DC structure is used to construct a specific rhombohedral lattice with axial angles equaling to $60°$. To obtain this lattice, one pair of the collinear creases remains unfolded.

Based on these three types of stacked origami structures, the 14 types of 3D Bravais lattices can be constructed according to Fig. S4. For each 3D Bravais lattice, a unit cell is shown on its corresponding origami structure. The detailed correlations between the origami geometries and lattice geometries are listed in Table S2.

**A  Triclinic: primitive**

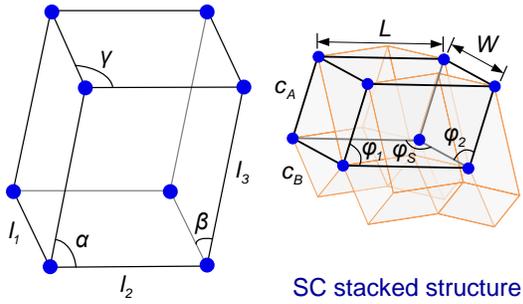

SC stacked structure

**B  (1) Monoclinic: primitive**

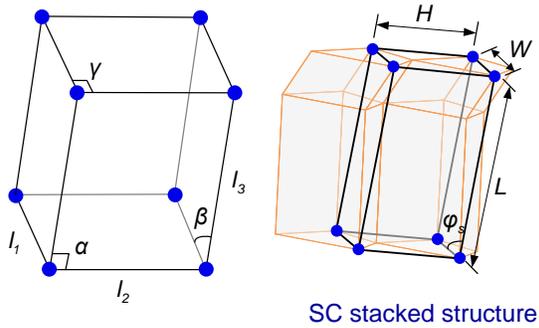

SC stacked structure

**(2) Monoclinic: based-centered**

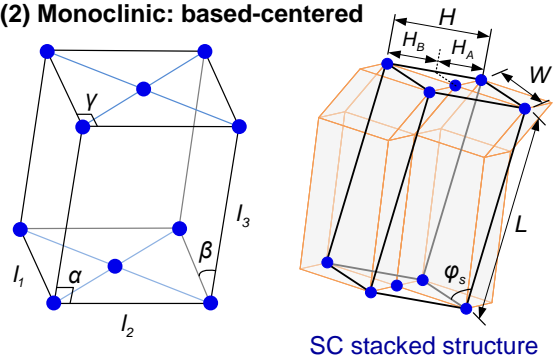

SC stacked structure

**C  (1) Orthorhombic: primitive**

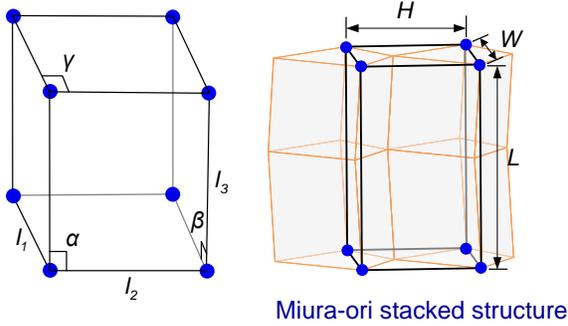

Miura-ori stacked structure

**(2) Orthorhombic: base-centered**

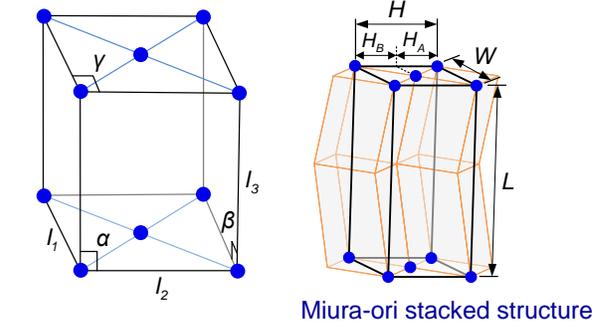

Miura-ori stacked structure

**(3) Orthorhombic: body-centered**

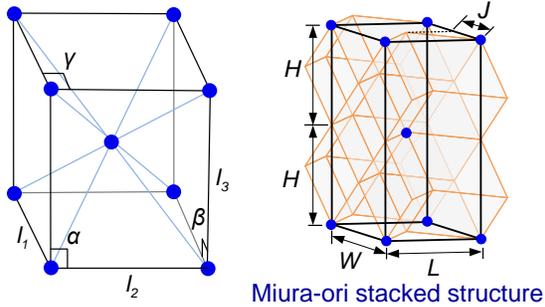

Miura-ori stacked structure

**(4) Orthorhombic: face-centered**

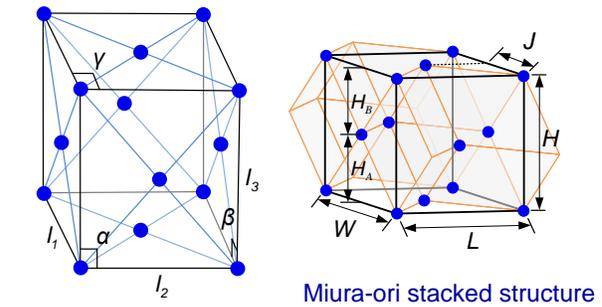

Miura-ori stacked structure

**Fig. S4. The 14 types of 3D Bravais lattices and the layout of a lattice unit cell on the corresponding origami structure**. (**A**) triclinic lattice (primitive), (**B**) monoclinic lattices (primitive and base-centered), (**C**) orthorhombic lattices (primitive, base-centered, body-centered, and face-centered).

**D** **(1) Tetragonal: primitive**

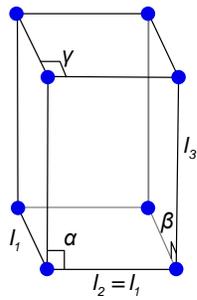

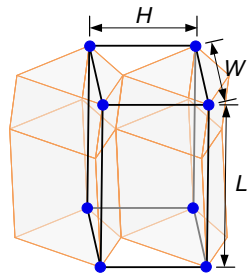

Miura-ori stacked structure

**(2) Tetragonal: base-centered**

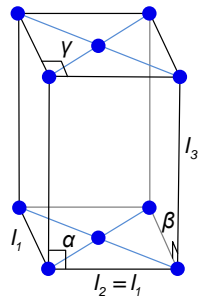

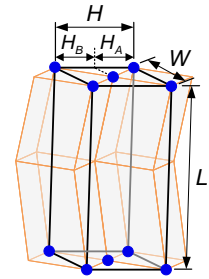

Miura-ori stacked structure

**E** **Rhombohedral: primitive**

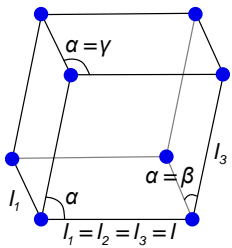

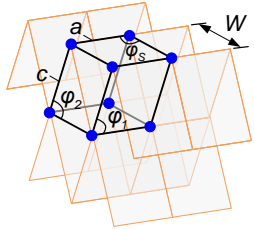

DC stacked structure

**F** **Hexagonal: primitive**

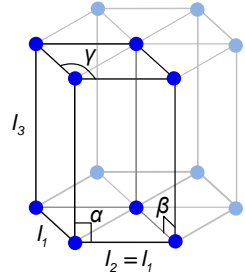

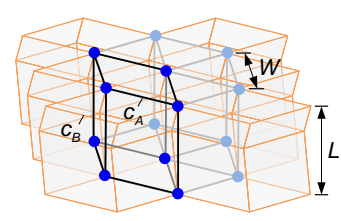

Miura-ori stacked structure

**G** **(1) Cubic: primitive**

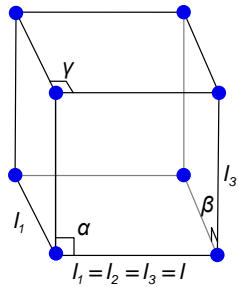

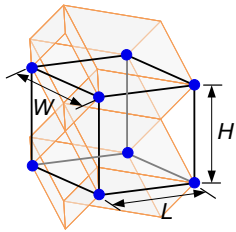

Miura-ori stacked structure

**(2) Cubic: body-centered**

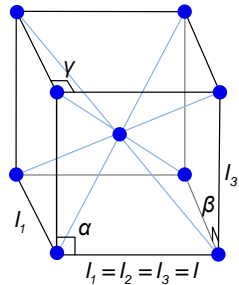

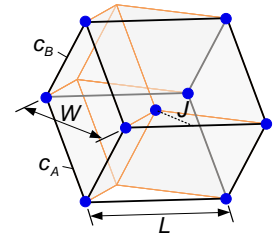

Miura-ori stacked structure

**(3) Cubic: face-centered**

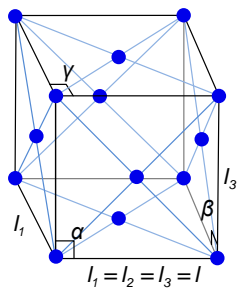

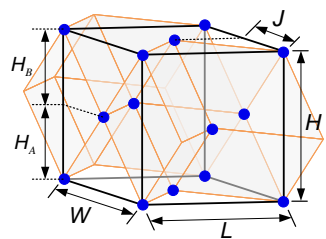

Miura-ori stacked structure

**Fig. S4 (continued). The 14 types of 3D Bravais lattices and the layout of a lattice unit cell on the corresponding origami structure**. (**D**) tetragonal lattices (primitive and base-centered), (**E**) rhombohedral lattice (primitive), (**F**) hexagonal lattice (primitive), (**G**) cubic lattices (primitive, base-centered, and face-centered).

**Table S2.** Geometry correlations for constructing 3D Bravais lattices based on 4-vertex origamis

| 14 types of 3D Bravais lattices | | | Axial distances and axial angles | Origami patterns | Correlation between lattice geometry and origami geometry |
|---|---|---|---|---|---|
| Lattice systems | Symmetry (Schönflies notations) | Centering types | | | |
| Triclinic | $C_1$ | Primitive | $l_1 \neq l_2 \neq l_3$, $\alpha \neq \beta \neq \gamma \neq 90°$ | Stacked SC structure | $W = l_1$, $L = l_2$, $c_A = l_3$ $\varphi_1 = \alpha$, $\varphi_2 = \beta$, $\varphi_S = \gamma$ |
| Monoclinic | $C_{2h}$ | Primitive | $l_1 \neq l_3$, $\beta \neq 90°$, $\alpha = \gamma = 90°$ | Stacked Miura-ori structure | $W = l_1$, $H = l_2$, $L = l_3$, $\varphi_S = \beta$ |
| | | Base-centered | | | $W = l_1$, $H = l_2$, $L = l_3$, $\varphi_S = \beta$, $H_A = H_B = H/2$ |
| Orthorhombic | $D_{2h}$ | Primitive | $l_1 \neq l_2 \neq l_3$, $\alpha = \beta = \gamma = 90°$ | Stacked Miura-ori structure | $W = l_1$, $H = l_2$, $L = l_3$ |
| | | Base-centered | | | $W = l_1$, $H = l_2$, $L = l_3$, $H_A = H_B = H/2$ |
| | | Body-centered | | | $W = l_1$, $L = l_2$, $H = l_3/2$, $J = W/2$ |
| | | Face-centered | | | $W = l_1$, $L = l_2$, $H = l_3$, $J = W/2$, $H_A = H_B = H/2$ |
| Tetragonal | $D_{4h}$ | Primitive | $l_1 = l_2 \neq l_3$, $\alpha = \beta = \gamma = 90°$ | Stacked Miura-ori structure | $W = H = l_1$, $L = l_3$ |
| | | Base-centered | | | $W = H = l_1$, $L = l_3$, $H_A = H_B = H/2$ |
| Rhombohedral | $D_{3d}$ | Primitive | $l_1 = l_2 = l_3 = l$, $\alpha = \beta = \gamma \neq 90°$ | Stacked DC structure | $a = c = W = l$, $\varphi_S = \varphi_1 = \varphi_2 = \gamma = 60°$ |
| Hexagonal | $D_{6h}$ | Primitive | $l_1 = l_2$, $\gamma = 120°$, $\alpha = \beta = 90°$ | Stacked Miura-ori structure | $c_A = c_B = W = l_1$, $L = l_3$ |
| Cubic | $O_h$ | Primitive | $l_1 = l_2 = l_3 = l$, $\alpha = \beta = \gamma = 90°$ | Stacked Miura-ori structure | $W = H = L = l$ |
| | | Body-centered | | | $c_A = c_B = L = l$, $J = W/2$, $W = \sqrt{2}L$ |
| | | Face-centered | | | $W = H = L = l$, $J = W/2$, $H_A = H_B = H/2$ |

# S3. Achieving diffusionless phase transformations through origami folding

Diffusionless phase transformations can be distinguished between transformations dominated by lattice-distortive strains and those where shuffles play a significant role. Lattice-distortive strains can transform the lattice from one Bravais type to another, and the shuffle refers to small movements of a lattice point within the lattice unit cell. In this section, we thoroughly examine the examples given in Fig. 2 (additional details of these examples are provided in Table S3). We show that different origami-based lattices can exhibit all kinds of 2D or 3D diffusionless transformation by rigid-folding. For each type of transformation, we formulate a lattice transformation matrix and correlate it to the folding kinematics. Such correlations are summarized in Table S4.

**Table S3.** Detailed geometries of origami patterns and origami folding showing in Fig.2

| Origami patterns used in Fig. 2 | | Folding angle | | Transformation type | |
|---|---|---|---|---|---|
| Pattern name | Parameters | Angle | Values used in Fig. 2 | 2D/3D | Type |
| Miura-ori sheet | $a=c,\ \gamma=60°$ | $\theta$ | $60° \to 30°$ | 2D | Dilation |
| Eggbox sheet | $a=b,\ \gamma=60°$ | $\alpha$ | $50° \to 45°$ | | Contraction/extension |
| SC sheet | $a=b=c,\ \gamma_1=36°,\ \gamma_2=72°$ | $\theta$ | $15° \to 65°$ | | With shear |
| Miura-ori sheet | $c/a=0.7,\ \gamma=60°$ | $\theta$ | $39.2° \to 50°$ | | With shuffle |
| Stacked GFF structure (bulged-out) | $a_A=c_A,\ \gamma_{B1}=54°,$ $\gamma_{A1}=36°,\ \gamma_{A2}=72°,$ | $\rho_{A1}$ | $10° \to 40°$ | 3D | Dilation |
| Stacked Miura-ori structure (bulged-out) | $a_A=c_A,\ \gamma_A=60°,\ \gamma_B=75°$ | $\theta_A$ | $-20° \to -60°$ | | Contraction/extension |
| GFF sheet | $a=c,\ \gamma_1=36°,\ \gamma_2=72°$ | $\rho_1$ | $90° \to 30°$ | | With shear |
| Stacked Miura-ori structure (nested-in) | $a_A=c_A,\ \gamma_A=60°,\ \gamma_B=75°$ | $\theta_A$ | $20° \to 60°$ | | With shuffle |

**Table S4.** Correlations between diffusionless lattice transformations and origami kinematic properties

| 2D/3D | Diffusionless lattice transformations | Origami kinematic properties | Representative origami structures |
|-------|---------------------------------------|------------------------------|-----------------------------------|
| 2D | Dilation | Negative in-plane Poisson's ratio | Miura-ori sheet |
| | Contraction/extension | Positive in-plane Poisson's ratio | Egg-box sheet |
| | Shear | In-plane shearing deformation mechanism | SC origami sheet |
| | Shuffle | Changes in the relative positions among characteristic entities | Miura-ori sheet |
| 3D | Dilation | Tri-directional auxetic effect (Negative Poisson's ratios in three directions) | Stacked Miura-ori structure (bulged-out) |
| | Contraction/extension | Opposite Poisson's ratios | GFF sheet |
| | Shear | Out-of-plane shearing deformation mechanism | Stacked Miura-ori structure (nested-in) |
| | Shuffle | Changes in the relative positions among characteristic entities | Stacked Miura-ori structure (bulged-out) |

## (1) 2D dilation

2D dilation is observed in a lattice of vertex inclusions based on Miura-ori, which is an example of 2D Bravais lattice shown in Fig. 2A and Fig. S5A. Since the folding of Miura-ori unit is described by the dihedral folding angle $\theta$ ($\theta \in [0, 90°]$) between its facet and the reference $x$-$o$-$y$ plane (Fig. S1A), we can formulate the lattice vectors corresponding to two different folding angles $\theta$ and $\theta'$:

$$\mathbf{l}_1^{\theta} = \begin{pmatrix} L|_{\theta} \\ 0 \end{pmatrix}, \mathbf{l}_2^{\theta} = \begin{pmatrix} 0 \\ W|_{\theta} \end{pmatrix}; \quad \mathbf{l}_1^{\theta'} = \begin{pmatrix} L|_{\theta'} \\ 0 \end{pmatrix}, \mathbf{l}_2^{\theta'} = \begin{pmatrix} 0 \\ W|_{\theta'} \end{pmatrix}. \tag{S11}$$

Here, the length $L$ and width $W$ of the Miura-ori unit are defined in Eq. (S2). Since there is no diffusion involved, we can formulate a *transformation matrix* $\mathbf{U}_{\text{2D-D}}$ to *quantitatively* describe the lattice transformation due to folding from angle $\theta$ to $\theta'$ such that

$$\left( \mathbf{l}_1^{\theta'} \quad \mathbf{l}_2^{\theta'} \right)^T = \mathbf{U}_{\text{2D-D}} \left( \mathbf{l}_1^{\theta} \quad \mathbf{l}_2^{\theta} \right)^T. \tag{S12}$$

$\mathbf{U}_{\text{2D-D}}$ describes the lattice-distortive strains that transform the lattice, and is also called the *Bain matrix*, in which the subscript "2D" indicates that this transformation is two-dimensional, and "D" means dilation. Substituting the expressions of $L$ and $W$ into Eq. (S11) and (S12), the transformation matrix can be written as

$$\mathbf{U}_{\text{2D-D}} = \begin{pmatrix} s \cdot \cos\theta'/\cos\theta & 0 \\ 0 & 1/s \end{pmatrix}, \text{ where } s = \frac{\sqrt{1-\sin^2\gamma\sin^2\theta}}{\sqrt{1-\sin^2\gamma\sin^2\theta'}}. \tag{S13}$$

$\mathbf{U}_{\text{2D-D}}$ is a diagonal matrix. When $\theta' < \theta$, the two diagonal elements of $\mathbf{U}_{\text{2D-D}}$ are always larger than 1, suggesting that the lattice dilates along both lattice vectors.

Note that the Miura-ori unit exhibits a negative Poisson's ratio in the $L$ and $W$ direction during folding (2)

$$v_{WL} = -\frac{\varepsilon_W}{\varepsilon_L} = -\frac{\mathrm{d}W/W}{\mathrm{d}L/L} = -\cos^2\theta\tan^2\gamma < 0. \tag{S14}$$

When the length $L$ increases (or decreases) due to folding, the width $W$ increases (or decrease) as well. Therefore, the negative Poisson's ratio is the kinematic origin of the 2D dilation.

## (2) 2D contraction/extension

2D contraction/extension is observed when transforming an eggbox-pattern-based lattice of vertex inclusions, which is an example of 2D Bravais lattice shown in Fig. 2B and Fig. S5B. The eggbox pattern (4) consists of four identical parallelogram facets characterized by crease lengths $a$, $b$ and a sector angle $\gamma$. The length $L$ and width $W$ of the eggbox unit are given by

$$W = 2a\sin\alpha, \ \ L = 2b\sin\beta, \tag{S15}$$

where $\alpha$ and $\beta$ are two angles between the boundary creases and a perpendicular line, used for describing the folding motion ($0 \leq \alpha, \beta \leq \gamma$) (Fig. S5B). They are not independent to each other but instead follow the relationship $\cos\alpha\cos\beta = \cos\gamma$. Hence, the lattice vectors corresponding to angle $\alpha$ and $\alpha'$ can be expressed as

$$\mathbf{l}_1^{\alpha} = \begin{pmatrix} L|_{\alpha} \\ 0 \end{pmatrix}, \mathbf{l}_2^{\alpha} = \begin{pmatrix} 0 \\ W|_{\alpha} \end{pmatrix}; \quad \mathbf{l}_1^{\alpha'} = \begin{pmatrix} L|_{\alpha'} \\ 0 \end{pmatrix}, \mathbf{l}_2^{\alpha'} = \begin{pmatrix} 0 \\ W|_{\alpha'} \end{pmatrix}. \tag{S16}$$

Since there is no diffusion, we can describe the lattice transformation due to folding from angle $\alpha$ to $\alpha'$ by formulating a transformation matrix $\mathbf{U}_{\text{2D-C/E}}$ such that

$$\begin{pmatrix} \mathbf{l}_1^{\alpha'} & \mathbf{l}_2^{\alpha'} \end{pmatrix}^T = \mathbf{U}_{\text{2D-C/E}} \begin{pmatrix} \mathbf{l}_1^{\alpha} & \mathbf{l}_2^{\alpha} \end{pmatrix}^T. \tag{S17}$$

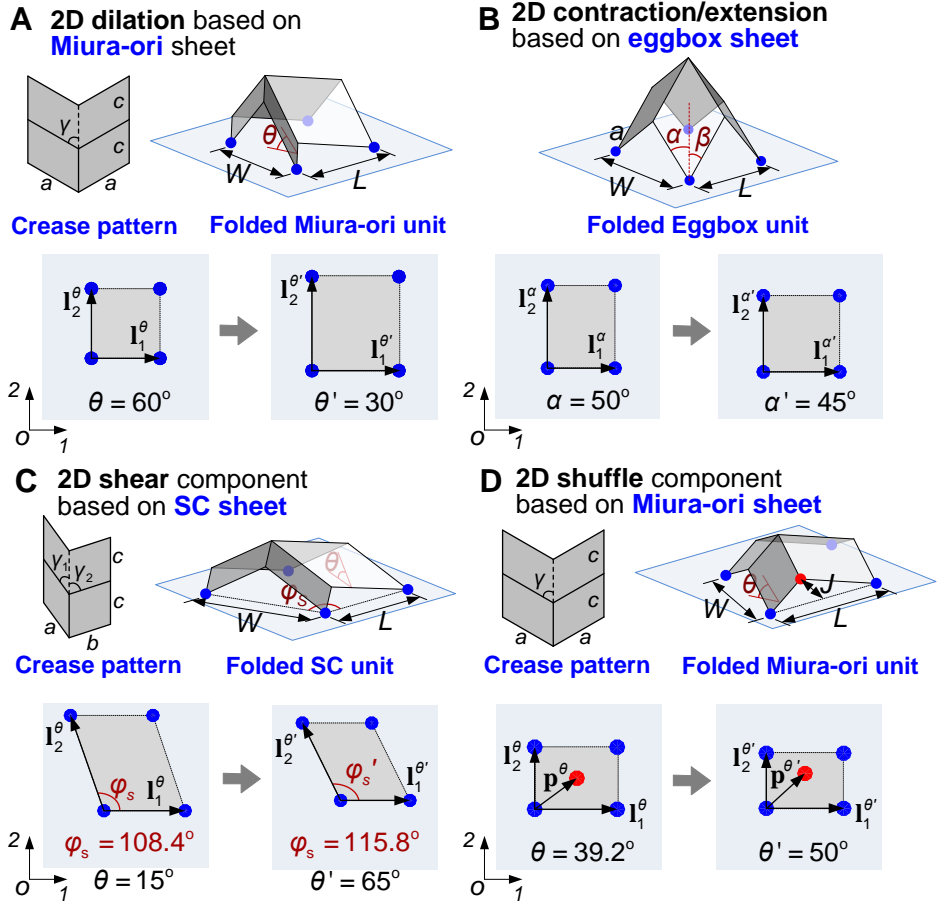

**Fig. S5. Origami geometries and 2D diffusionless phase transformations**. (**A**) 2D dilation achieved by transforming a Miura-ori-based lattice of vertex inclusions; (**B**) 2D contraction/extension achieved by transforming an eggbox-pattern-based lattice of vertex inclusions; (**C**) 2D shear involved in transforming an SC-ori-based lattice of vertex inclusions; (**D**) 2D shuffle involved in transforming a Miura-ori based lattice of vertex inclusions.

Here $\mathbf{U}_{\text{2D-C/E}}$ describes the lattice-distortive strains that transform the lattice; the subscript "2D-C/E" stands for two-dimensional contraction and extension. Substituting the expressions of $L$ and $W$ into Eq. (S16) and (S17), the transformation matrix becomes

$$\mathbf{U}_{\text{2D-C/E}} = \begin{pmatrix} \sin\beta'/\sin\beta & 0 \\ 0 & \sin\alpha'/\sin\alpha \end{pmatrix}. \tag{S18}$$

$\mathbf{U}_{\text{2D-C/E}}$ is also a diagonal matrix. When $\alpha' > \alpha$, $\sin\alpha'/\sin\alpha > 1$ but $0 < \sin\beta'/\sin\beta < 1$, suggesting that the lattice contracts along one lattice vector but expands along the other vector, and vice versa.

Note that the eggbox unit exhibits a positive Poisson's ratio in the $L$ and $W$ directions during folding [4]

$$v_{WL} = -\frac{\varepsilon_W}{\varepsilon_L} = -\frac{\mathrm{d}W/W}{\mathrm{d}L/L} = \frac{\cos^2 \alpha - \cos^2 \gamma}{\cos^2 \theta \tan^2 \gamma} > 0. \tag{S19}$$

When the length $L$ increases due to folding, the width $W$ decreases instead; and vice versa. Therefore, the positive Poisson's ratio is kinematic origin of the 2D contraction/extension.

**(3) 2D shear**

2D shear component is observed when we transform a lattice of vertex inclusions based on single collinear (SC) origami, which is an example of 2D Bravais lattice shown in Fig. 2C and Fig. S5C. Folding of the SC unit can be described by the dihedral angle $\theta$ ($\theta \in [0, 90^\circ]$) defined between the facet and the reference *x-o-y* plane (Fig. S1A) [1], so that the lattice vectors corresponding to angle $\theta$ and $\theta'$ can be expressed as

$$\mathbf{l}_1^\theta = \begin{pmatrix} L|_\theta \\ 0 \end{pmatrix}, \ \mathbf{l}_2^\theta = \begin{pmatrix} (W\cos\varphi_S)|_\theta \\ (W\sin\varphi_S)|_\theta \end{pmatrix}; \quad \mathbf{l}_1^{\theta'} = \begin{pmatrix} L|_{\theta'} \\ 0 \end{pmatrix}, \ \mathbf{l}_2^{\theta'} = \begin{pmatrix} (W\cos\varphi_S)|_{\theta'} \\ (W\sin\varphi_S)|_{\theta'} \end{pmatrix}. \tag{S20}$$

The length $L$, width $W$, and the angle $\varphi_S$ of the SC unit are defined in Eq. (S1). Since there is no diffusion, we can describe the lattice transformation due to folding from angle $\theta$ to $\theta'$ based on a transformation matrix $\mathbf{U}_{\text{2D-S}}$ such that

$$\begin{pmatrix} \mathbf{l}_1^{\theta'} & \mathbf{l}_2^{\theta'} \end{pmatrix}^T = \mathbf{U}_{\text{2D-S}} \begin{pmatrix} \mathbf{l}_1^\theta & \mathbf{l}_2^\theta \end{pmatrix}^T. \tag{S21}$$

Here $\mathbf{U}_{\text{2D-S}}$ describes the lattice-distortive strains that transform the lattice; the subscript "2D-S" means two-dimensional shear. $\mathbf{U}_{\text{2D-S}}$ can be formulated as:

$$\mathbf{U}_{\text{2D-S}} = \begin{pmatrix} L|_{\theta'}/L|_\theta & 0 \\ \dfrac{W|_{\theta'} \cdot \left(\cos\varphi_S|_{\theta'} - \sin\varphi_S|_{\theta'}/\tan\varphi_S|_\theta\right)}{L|_\theta} & \dfrac{(W\sin\varphi_S)|_{\theta'}}{(W\sin\varphi_S)|_\theta} \end{pmatrix}. \tag{S22}$$

One can clearly see that $\mathbf{U}_{\text{2D-S}}$ is not a diagonal matrix. That is, in addition to the non-trivial diagonal elements that describes the contraction or extension along the lattice vectors, there is also a non-zero off-diagonal element that indicates a distortion by shearing. By substituting the expressions of $L$, $W$, and $\varphi_S$ into Eq. (S22), one can connect $\mathbf{U}_{\text{2D-S}}$ to the rigid-folding kinematics.

Note that the SC origami is characterized by its ability to achieve in-plane shearing deformation during folding, which is manifested by the changes in angle $\varphi_S$. For example, by folding the SC unit ($a = b = c$, $\gamma_1 = 36°$, $\gamma_2 = 72°$) from $\theta = 15°$ to $\theta = 65°$, the angle $\varphi_S$ changes from $108.4°$ to $115.8°$. As a result, the in-plane shearing deformation mechanism is the origin of 2D shear during lattice transformation.

**(4) 2D shuffle**

2D shuffle plays a significant role in transforming a particular type of Miura-ori based lattice of vertex inclusions shown in Fig. 2D and Fig. S5D. This lattice as a whole is not a Bravais type due to the additional lattice point within the unit cell. As a result, the two lattice vectors used in the previous three cases are no longer sufficient to describe the lattice, because the lattice point inside the unit cell cannot be obtained by translating the two vectors. This kind of lattice is essentially a collection of two congruent Bravais lattices that are shifted from one another. Mathematically, such *multi-lattice* (5) can be characterized by a group of lattice vectors $\mathbf{l}_i$ and a *shift* vector $\mathbf{p}$. That is, $\mathbf{l}_i$ describe the constituent Bravais lattice, and $\mathbf{p}$ describes the offset between the two congruent lattices. Here, the folding of Miura-ori unit is still described by the dihedral angle $\theta$ ($\theta \in [0,90°]$) between its facet and the reference *x-o-y* plane (Fig. S1A) (1), so that the vectors corresponding to angle $\theta$ and $\theta'$ can be expressed as

$$\mathbf{l}_1^\theta = \begin{pmatrix} L|_\theta \\ 0 \end{pmatrix}, \quad \mathbf{l}_2^\theta = \begin{pmatrix} 0 \\ W|_\theta \end{pmatrix}, \quad \mathbf{p}^\theta = \begin{pmatrix} L|_\theta / 2 \\ J|_\theta \end{pmatrix}; \quad \mathbf{l}_1^{\theta'} = \begin{pmatrix} L|_{\theta'} \\ 0 \end{pmatrix}, \quad \mathbf{l}_2^{\theta'} = \begin{pmatrix} 0 \\ W|_{\theta'} \end{pmatrix}, \quad \mathbf{p}^{\theta'} = \begin{pmatrix} L|_{\theta'} / 2 \\ J|_{\theta'} \end{pmatrix}. \quad (S23)$$

The dimensions $L$, $W$, and $J$ of the Miura-ori unit are defined in Eq. (S2). Since there is no diffusion, we can then describe the lattice transformation induced by folding from angle $\theta$ to $\theta'$ as a combination of deformation and shift. Therefore, we can formulate the transformation matrix $\mathbf{T}_{\text{2D-SH}}$ such that

$$\left( \mathbf{l}_1^{\theta'} \quad \mathbf{l}_2^{\theta'} \quad \mathbf{p}^{\theta'} \right)^T = \mathbf{T}_{\text{2D-SH}} \left( \mathbf{l}_1^{\theta} \quad \mathbf{l}_2^{\theta} \quad \mathbf{p}^{\theta} \right)^T. \quad (S24)$$

Here the subscript "2D-SH" means two-dimensional shuffle, and $\mathbf{T}_{\text{2D-SH}}$ can be written as

$$\mathbf{T}_{\text{2D-SH}} = \begin{pmatrix} & & 0 \\ \mathbf{U}_{\text{2D-D}} & & 0 \\ \mu_1 & \mu_2 & \lambda \end{pmatrix}. \quad (S25)$$

$\mathbf{T}_{2D\text{-}SH}$ is not a diagonal matrix, and it reflects how the lattice-distortive strain and shuffle component co-exist in the transformation. $\mathbf{U}_{2D\text{-}D}$ is the submatrix corresponding to 2D dilation, which describes the lattice-distortive strains of the constituent 2D Bravias lattice. It can be replaced by other transformation matrices (i.e., $\mathbf{U}_{2D\text{-}D}$, $\mathbf{U}_{2D\text{-}C/E}$, or $\mathbf{U}_{2D\text{-}S}$) if different origami structures are employed. Other elements in the third row of $\mathbf{T}_{2D\text{-}SH}$ quantify the 2D shuffle component, which satisfy the following relationship

$$\left(\mathbf{p}^{\theta'}\right)^T = \begin{pmatrix} \mu_1 & \mu_2 & \lambda \end{pmatrix}\begin{pmatrix} \mathbf{l}_1^\theta & \mathbf{l}_2^\theta & \mathbf{p}^\theta \end{pmatrix}^T. \tag{S26}$$

By substituting the expressions of $L$, $W$, and $J$ into Eq. (S23) and Eq. (S24), one can connect $\mathbf{T}_{2D\text{-}SH}$ to the rigid-folding kinematics.

Note that changes in the relative positions among vertices (or other characteristic entities) is ubiquitous for origami folding, and it is the underlying mechanism that generates the shuffle during lattice transformation. The shift component can be significant, and it can even move a lattice point from one unit cell to another, triggering a break and reconstruction of the nearest-neighbor relationship (e.g., Fig. 4B in the main text).

**(5) 3D dilation**

3D dilation is observed when transforming a lattice of vertex inclusions based on the stacked generic-flat-foldable (GFF) structure in a bulged-out configuration (1) (a kind of 3D Bravias lattice shown in Fig. S6A). The stacked GFF structure consists of two GFF units A and B. Each GFF unit can be characterized by crease lengths $a_x$, $c_x$, and sector angles $\gamma_{x1}$, $\gamma_{x2}$. The subscript '$x$' takes the values of '$A$' or '$B$' and it stands for the bottom unit A or the top unit B (Fig. S6A, left). Without the loss of generality, we assume $\gamma_{x1} < \gamma_{x2}$. The following geometry constraints have to be reinforced to ensure the kinematic compatibility between two GFF units

$$a_A = a_B, \quad \frac{\cos\gamma_{A1}}{\cos\gamma_{B1}} = \frac{\cos\gamma_{A2}}{\cos\gamma_{B2}} = \frac{c_B}{c_A}. \tag{S27}$$

Folding of the stacked GFF structure is still a single degree-of-freedom mechanism, and it can be described by the dihedral angle $\rho_{A1}$ between two facets of the bottom unit A (Fig. S6A, right). Alternatively, we can describe its folding by using the dihedral angles $\theta_{Ai}$ (or $\theta_{Bi}$) ($i = 1, 2, 3, 4$) between the facets of the bottom unit A (or top unit B) and the reference $x$-$o$-$y$ plane. $\theta_{Ai}$ and $\theta_{Bi}$ are not independent to each other but satisfy the following constraints

$$
\begin{aligned}
&\cos\theta_{A1}\tan\gamma_{A1}=\cos\theta_{B1}\tan\gamma_{B1},\\
&\sin\theta_{A1}\sin\gamma_{A1}=\sin\theta_{A2}\sin\gamma_{A2}=\sin\theta_{A3}\sin\gamma_{A1}=\sin\theta_{A4}\sin\gamma_{A2},\\
&\sin\theta_{B1}\sin\gamma_{B1}=\sin\theta_{B2}\sin\gamma_{B2}=\sin\theta_{B3}\sin\gamma_{B1}=\sin\theta_{B4}\sin\gamma_{B2}.
\end{aligned}
\tag{S28}
$$

When $\rho_{A1}^{C1}\le\rho_{A1}<180^{\circ}$, the structure is in nested-in configurations and $0^{\circ}<\theta_{Ai}\le90^{\circ}$. In particular, when $\rho_{A1}=\rho_{A1}^{C1}$, the structure self-locks so that $\theta_{A1}=\theta_{A3}=90^{\circ}$ and $0^{\circ}<\theta_{A2}=\theta_{A4}<90^{\circ}$. On the other hand, when $180^{\circ}\le\rho_{A1}\le360^{\circ}$, the structure is in bulged-out configurations and $-180^{\circ}\le\theta_{Ai}\le0^{\circ}$.

Here we focus on the bulged-out configuration of the stacked GFF structure. For $180^{\circ}\le\rho_{A1}\le360^{\circ}$, the folding can be divided into two stages. In the first stage, $\rho_{A1}$ increases from $180^{\circ}$ to a critical value $\rho_{A1}^{C2}$ ($\rho_{A1}^{C2}=\rho_{A1}^{C1}+90^{\circ}$), and $|\theta_{Ai}|$ increases accordingly from $0^{\circ}$ but remains smaller than $90^{\circ}$. The first stage ends when $\rho_{A1}$ reaches a critical value $\rho_{A1}^{C2}$. At this critical configuration, both $|\theta_{A1}|$ and $|\theta_{A3}|$ reach $90^{\circ}$ prior to $|\theta_{A2}|$ and $|\theta_{A4}|$. In the second stage of folding, $\rho_{A1}$ continues to increase beyond $\rho_{A1}^{C2}$ ($\rho_{A1}^{C2}<\rho_{A1}\le360^{\circ}$), $|\theta_{A1}|$ and $|\theta_{A3}|$ also continue increasing so that $|\theta_{A1}|=|\theta_{A3}|>90^{\circ}$, however, $|\theta_{A2}|$ and $|\theta_{A4}|$ decrease and remain smaller than $90^{\circ}$ ($|\theta_{A2}|=|\theta_{A4}|<90^{\circ}$). Details of the folding kinematics are discussed in (*24*). Specifically, $\theta_{Ai}$ ($i=1,2,3,4$) can be expressed as

$$
\theta_{A1}=\theta_{A3}=\begin{cases}\arcsin\dfrac{\sin\gamma_{A2}\sin\rho_{A1}}{\sqrt{\sin^2\gamma_{A1}+\sin^2\gamma_{A2}-2\sin\gamma_{A1}\sin\gamma_{A2}\cos\rho_{A1}}}, & 180^{\circ}\le\rho_{A1}\le\rho_{A1}^{C2},\\[3mm]
-\arcsin\dfrac{\sin\gamma_{A2}\sin\rho_{A1}}{\sqrt{\sin^2\gamma_{A1}+\sin^2\gamma_{A2}-2\sin\gamma_{A1}\sin\gamma_{A2}\cos\rho_{A1}}}-180^{\circ}, & \rho_{A1}^{C2}<\rho_{A1}\le360^{\circ},
\end{cases}
\tag{S29}
$$

$$
\theta_{A2}=\theta_{A4}=\arcsin\dfrac{\sin\gamma_{A1}\sin\rho_{A1}}{\sqrt{\sin^2\gamma_{A1}+\sin^2\gamma_{A2}-2\sin\gamma_{A1}\sin\gamma_{A2}\cos\rho_{A1}}}, \qquad 180^{\circ}\le\rho_{A1}\le360^{\circ}.
$$

At the bulged-out configuration, the outer dimensions (length $L$, width $W$, and height $H$) of the stacked GFF structure are (Fig. S6A, right):

$$
\begin{aligned}
&L=\begin{cases}2a_A\sin\big((\xi+\eta)/2\big), & 180^{\circ}\le\rho_{A1}\le\rho_{A1}^{C2},\\
2a_A\sin\big((-\xi+\eta)/2\big), & \rho_{A1}^{C2}<\rho_{A1}\le360^{\circ},\end{cases}\\[2mm]
&W=\begin{cases}2c_A\sqrt{1-\sin^2\gamma_{A1}\sin^2\theta_{A1}}\,\cos\big((\xi-\eta)/2\big), & 180^{\circ}\le\rho_{A1}\le\rho_{A1}^{C2},\\
2c_A\sqrt{1-\sin^2\gamma_{A1}\sin^2\theta_{A1}}\,\cos\big((\xi+\eta)/2\big), & \rho_{A1}^{C2}<\rho_{A1}\le360^{\circ},\end{cases}\\[2mm]
&H_A=c_A\sin\gamma_{A1}\sin|\theta_{A1}|,\ H_B=c_B\sin\gamma_{B1}\sin\theta_{B1},\ H=H_A+H_B,\ \ 180^{\circ}\le\rho_{A1}\le360^{\circ},
\end{aligned}
\tag{S30}
$$

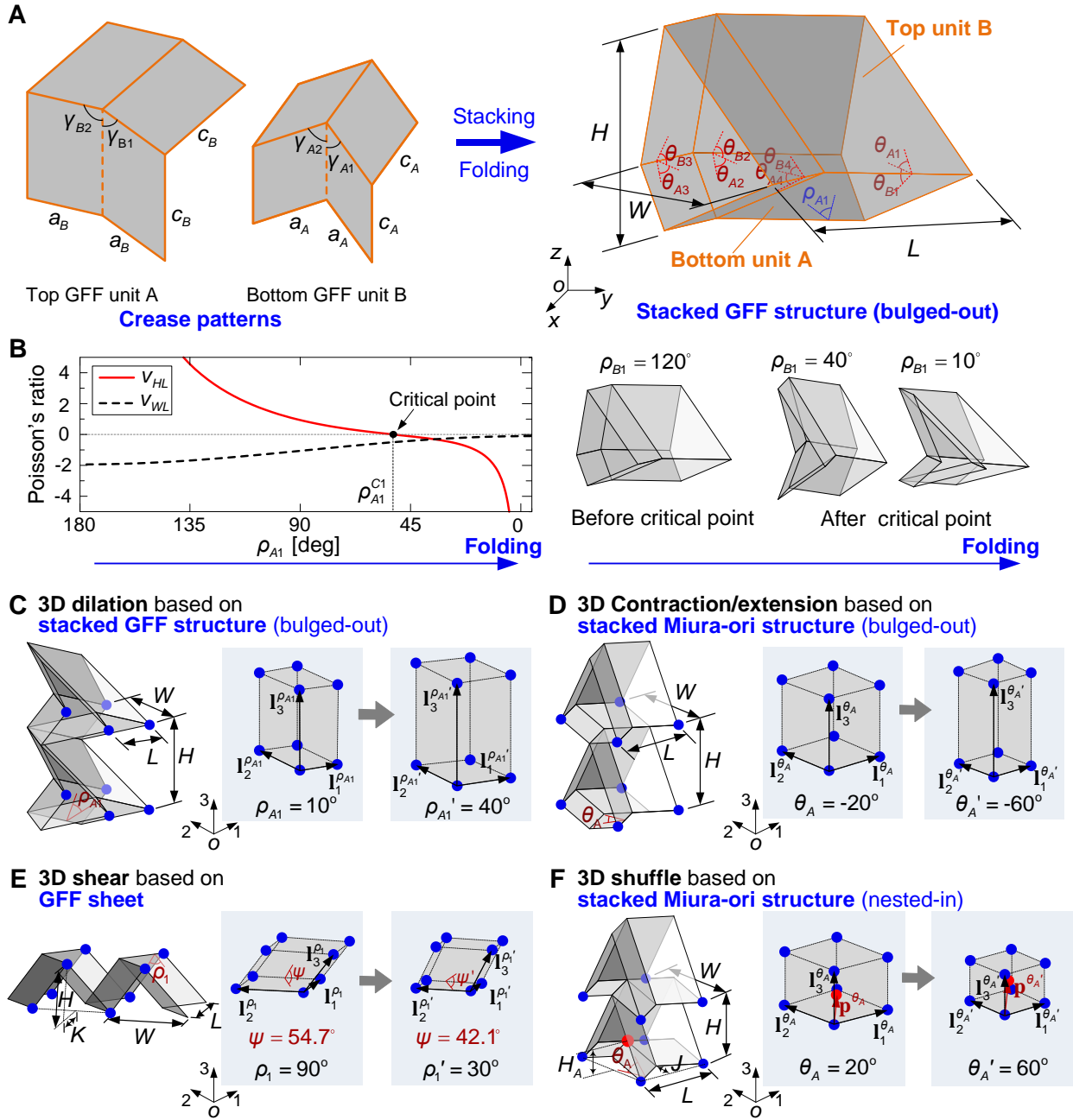

**Fig. S6. Origami geometries and 3D diffusionless phase transformations**. (**A**) Geometries of a stacked GFF structure (bulged-out), where the constituent units' crease patterns (left) and the origami at a folded state (right) are given. (**B**) Poisson's ratios of the stacked GFF structure (bulged-out). (**C**) 3D dilation achieved by transforming a lattice of vertex inclusions based on stacked GFF structure (bulged-out); (**D**) 3D contraction/extension achieved by transforming a lattice of vertex inclusions based on stacked Miura-ori structure (bulged-out); (**E**) 3D shear involved in transforming a lattice of vertex inclusions based on GFF sheet; (**F**) 3D shuffle involved in transforming a lattice of vertex inclusions based on stacked Miura-ori structure (nested-in).

where $\xi$ and $\eta$ are similarly defined as Eq. (S5) so that

$$\xi = \arccos \frac{\cos \gamma_{A1}}{\sqrt{1 - \sin^2 \gamma_{A1} \sin^2 \theta_{A1}}}, \quad \eta = \arccos \frac{\cos \gamma_{A2}}{\sqrt{1 - \sin^2 \gamma_{A1} \sin^2 \theta_{A1}}}. \tag{S31}$$

Based on the assignment of lattice points in the stacked GFF structure given in Fig. S6C, the lattice vectors corresponding to the folding angle $\rho_{A1}$ and $\rho_{A1}{}'$ can be formulated as

$$\mathbf{l}_1^{\rho_{A1}} = \begin{pmatrix} L|_{\rho_{A1}} \\ 0 \\ 0 \end{pmatrix}, \quad \mathbf{l}_2^{\rho_{A1}} = \begin{pmatrix} 0 \\ W|_{\rho_{A1}} \\ 0 \end{pmatrix}, \quad \mathbf{l}_3^{\rho_{A1}} = \begin{pmatrix} 0 \\ 0 \\ H|_{\rho_{A1}} \end{pmatrix};$$

$$\mathbf{l}_1^{\rho_{A1}{}'} = \begin{pmatrix} L|_{\rho_{A1}{}'} \\ 0 \\ 0 \end{pmatrix}, \quad \mathbf{l}_2^{\rho_{A1}{}'} = \begin{pmatrix} 0 \\ W|_{\rho_{A1}{}'} \\ 0 \end{pmatrix}, \quad \mathbf{l}_3^{\rho_{A1}{}'} = \begin{pmatrix} 0 \\ 0 \\ H|_{\rho_{A1}{}'} \end{pmatrix}. \tag{S32}$$

Since there is no diffusion, we can then describe the lattice transformation due to folding from angle $\rho_{A1}$ to $\rho_{A1}{}'$ based on a transformation matrix $\mathbf{U}_{\text{3D-D}}$ such that

$$\begin{pmatrix} \mathbf{l}_1^{\rho_{A1}{}'} & \mathbf{l}_2^{\rho_{A1}{}'} & \mathbf{l}_3^{\rho_{A1}{}'} \end{pmatrix}^T = \mathbf{U}_{\text{3D-D}} \begin{pmatrix} \mathbf{l}_1^{\rho_{A1}} & \mathbf{l}_2^{\rho_{A1}} & \mathbf{l}_3^{\rho_{A1}} \end{pmatrix}^T. \tag{S33}$$

Here $\mathbf{U}_{\text{3D-D}}$ is a diagonal matrix and it describes the lattice-distortive strains that transform the lattice; the subscript "3D-D" means three-dimensional dilation. $\mathbf{U}_{\text{3D-D}}$ can be written as

$$\mathbf{U}_{\text{3D-D}} = \begin{pmatrix} L|_{\rho_{A1}{}'} / L|_{\rho_{A1}} & 0 & 0 \\ 0 & W|_{\rho_{A1}{}'} / W|_{\rho_{A1}} & 0 \\ 0 & 0 & H|_{\rho_{A1}{}'} / H|_{\rho_{A1}} \end{pmatrix}. \tag{S34}$$

By substituting the expressions of $L$, $W$ and $H$ into Eq. (S34), one can correlate $\mathbf{U}_{\text{3D-D}}$ to the rigid-folding kinematics.

The observed 3D dilation can also be related to the auxetic properties of the stacked GFF structure. Based on its outer dimensions, the Poisson's ratios can be calculated via

$$v_{WL} = -\frac{\varepsilon_W}{\varepsilon_L} = -\frac{\mathrm{d}W / W}{\mathrm{d}L / L}, \quad v_{HL} = -\frac{\varepsilon_H}{\varepsilon_L} = -\frac{\mathrm{d}H / H}{\mathrm{d}L / L}, \quad v_{WH} = -v_{WL} / v_{HL}. \tag{S35}$$

Fig. S6B shows the values of the Poisson's ratios ($v_{WL}$ and $v_{HL}$) of the stacked GFF structure with respect to folding at the bulged-out configuration (i.e. $180^\circ \leq \rho_{A1} \leq 360^\circ$). After the critical point $\rho_{A1}^{C2}$, the stacked GFF structure exhibits negative Poisson's ratios in all the three directions. Therefore, in the folding range $\rho_{A1}^{C2} < \rho_{A1} \leq 360^\circ$, if $L|_{\rho_{A1}}' > L|_{\rho_{A1}}$, the negative Poisson's ratios indicate that $W_{\rho_{A1}'} > W|_{\rho_{A1}}$ and $H|_{\rho_{A1}'} > H|_{\rho_{A1}}$. This suggests that the lattice dilates along all of the three lattice vectors according to the transformation matrix $\mathbf{U}_{\text{3D-D}}$ in Eq. (S34). Therefore, the tri-directional auxetic effect (i.e., negative Poisson's ratios in all the three directions) is the kinematic origin of 3D dilation transformation.

### (6) 3D contraction/extension

3D contraction/extension is observed when transforming a 3D Bravais lattice of vertex inclusions based on the stacked Miura-ori structure in a bulged-out configuration ($-90^\circ < \theta_A < 0$) (1) , shown in Fig. S6D. Construction of the stacked Miura-ori structure is introduced in Fig. S3B, and its outer dimensions are given in Eq. (S9).

Based on the assignment of lattice points in the stacked Miura-ori structure given in Fig. S6D, the lattice vectors corresponding to folding angle $\theta_A$ and $\theta_A'$ can be expresses as

$$\mathbf{l}_1^{\theta_A} = \begin{pmatrix} L|_{\theta_A} \\ 0 \\ 0 \end{pmatrix}, \quad \mathbf{l}_2^{\theta_A} = \begin{pmatrix} 0 \\ W|_{\theta_A} \\ 0 \end{pmatrix}, \quad \mathbf{l}_3^{\theta_A} = \begin{pmatrix} 0 \\ 0 \\ H|_{\theta_A} \end{pmatrix};$$

$$\mathbf{l}_1^{\theta_A'} = \begin{pmatrix} L|_{\theta_A'} \\ 0 \\ 0 \end{pmatrix}, \quad \mathbf{l}_2^{\theta_A'} = \begin{pmatrix} 0 \\ W|_{\theta_A'} \\ 0 \end{pmatrix}, \quad \mathbf{l}_3^{\theta_A'} = \begin{pmatrix} 0 \\ 0 \\ H|_{\theta_A'} \end{pmatrix}.$$

(S36)

Since there is no diffusion, we can then describe the lattice transformation due to folding from angle $\theta_A$ to $\theta_A'$ based on a transformation matrix $\mathbf{U}_{\text{3D-C/E}}$ such that

$$\left( \mathbf{l}_1^{\theta_A'} \quad \mathbf{l}_2^{\theta_A'} \quad \mathbf{l}_3^{\theta_A'} \right)^T = \mathbf{U}_{\text{3D-D}} \left( \mathbf{l}_1^{\theta_A} \quad \mathbf{l}_2^{\theta_A} \quad \mathbf{l}_3^{\theta_A} \right)^T.$$

(S37)

Here $\mathbf{U}_{\text{3D-C/E}}$ is a diagonal matrix and it describes the lattice-distortive strains that transform the lattice; the subscript "3D-C/E" means three-dimensional contraction/expansion. $\mathbf{U}_{\text{3D-C/E}}$ can be formulated as

$$\mathbf{U}_{\text{3D-C/E}} = \begin{pmatrix} L|_{\theta_{A'}} / L|_{\theta_A} & 0 & 0 \\ 0 & W|_{\theta_{A'}} / W|_{\theta_A} & 0 \\ 0 & 0 & H|_{\theta_{A'}} / H|_{\theta_A} \end{pmatrix}. \tag{S38}$$

By substituting the expressions of $L$, $W$, and $H$ into Eq. (S38), one can correlate $\mathbf{U}_{\text{3D-C/E}}$ to the rigid-folding kinematics.

The observed 3D contraction/extension is also related to Poisson's ratios. Based on the outer dimensions of the stacked Miura-ori structure in bulged-out configurations ($-90° < \theta_A < 0$), the Poisson's ratios $v_{WL}$ and $v_{HL}$ have opposite signs in that

$$v_{WL} = -\frac{\varepsilon_W}{\varepsilon_L} = -\frac{\mathrm{d}W}{\mathrm{d}L}\frac{L}{W} = -\cos^2\theta_A \tan^2\gamma_A < 0,$$

$$v_{HL} = -\frac{\varepsilon_H}{\varepsilon_L} = -\frac{\mathrm{d}H}{\mathrm{d}L}\frac{L}{H} = -\frac{\tan\gamma_A\left(1 - \sin^2\gamma_A \sin^2\theta_A\right)}{\sin^2\gamma_A \sin\theta_A \sqrt{\tan^2\gamma_B - \cos^2\theta_A \tan^2\gamma_A}} > 0. \tag{S39}$$

Hence, in the folding range $-90° < \theta_A < 0$, if $L|_{\theta_{A'}} < L|_{\theta_A}$, the negative Poisson's ratios $v_{WL}$ indicates that $W_{\theta_{A'}} < W|_{\theta_A}$, however, the positive Poisson's ratio $v_{HL}$ indicates that $H|_{\theta_{A'}} > H|_{\theta_A}$. As a result, the lattice contracts in the $L$ and $W$ directions but extends in the $H$ direction. The opposite Poisson's ratios are the kinematic origin of 3D contraction/extension.

## (7) 3D shear

3D shear is observed when transforming a 3D Bravais lattice of vertex inclusions based on the GFF sheet (1), shown in Fig. S6E. A GFF unit is characterized by crease lengths $a, c$, and sector angles $\gamma_1, \gamma_2$. Without the loss of generality, we assume $\gamma_1 < \gamma_2$. Its folding can be described by the dihedral angle $\rho_1$ or the folding angles $\theta_i$ ($i = 1,2,3,4$), which are defined as the dihedral angle between the facets and the reference $x$-$o$-$y$ plane. Folding of a GFF unit can also be divided into two stages. In the first stage, $\rho_1$ decreases from $180°$ to a critical value $\rho_1^C$, while $\theta_i$ increase from $0°$ but remain smaller than $90°$. At the end of the first stage, $\rho_1$ reaches a critical value $\rho_1^C$, while $\theta_1$ and $\theta_3$ reach $90°$ prior to $\theta_2$ and $\theta_4$. In the second stage, $\rho_1$ continue to decrease below $\rho_1^C$ towards 0. Meanwhile, $\theta_1$ and $\theta_3$ keep increasing ($\theta_1 = \theta_3 > 90°$), but $\theta_2$ and $\theta_4$ start to decrease and remain $\theta_2 = \theta_4 < 90°$. Details of the folding kinematics are discussed in (1). Specifically, $\theta_i$ ($i = 1,2,3,4$) can be expressed as

$$\theta_1 = \theta_3 = \begin{cases} \arcsin \dfrac{\sin \gamma_2 \sin \rho_1}{\sqrt{\sin^2 \gamma_1 + \sin^2 \gamma_2 - 2\sin \gamma_1 \sin \gamma_2 \cos \rho_1}}, & \rho_1^C \le \rho_1 \le 180^\circ, \\[4mm] 180^\circ - \arcsin \dfrac{\sin \gamma_2 \sin \rho_1}{\sqrt{\sin^2 \gamma_1 + \sin^2 \gamma_2 - 2\sin \gamma_1 \sin \gamma_2 \cos \rho_1}}, & 0 \le \rho_1 < \rho_1^C, \end{cases} \quad \text{(S40)}$$

$$\theta_2 = \theta_4 = \arcsin \dfrac{\sin \gamma_1 \sin \rho_1}{\sqrt{\sin^2 \gamma_1 + \sin^2 \gamma_2 - 2\sin \gamma_1 \sin \gamma_2 \cos \rho_1}}, \quad 0 \le \rho_1 \le 180^\circ.$$

The outer dimensions of a GFF unit are

$$L = \begin{cases} 2a \sin\left((\xi + \eta)/2\right), & \rho_1^C \le \rho_1 \le 180^\circ, \\[2mm] 2a \sin\left((-\xi + \eta)/2\right), & 0 \le \rho_1 < \rho_1^C, \end{cases}$$

$$W = \begin{cases} 2c\sqrt{1 - \sin^2 \gamma_1 \sin^2 \theta_1}\, \cos\left((\xi - \eta)/2\right), & \rho_1^C \le \rho_1 \le 180^\circ, \\[2mm] 2c\sqrt{1 - \sin^2 \gamma_1 \sin^2 \theta_1}\, \cos\left((\xi + \eta)/2\right), & 0 \le \rho_1 < \rho_1^C, \end{cases} \quad \text{(S41)}$$

$$H = c \sin \gamma_1 \sin \theta_1, \quad K = \sqrt{a^2 - H^2 - \left(W/2\right)^2}, \quad 0 \le \rho_1 \le 180^\circ,$$

where $\xi$ and $\eta$ are similarly defined as Eq. (S1), i.e.,

$$\xi = \arccos \dfrac{\cos \gamma_1}{\sqrt{1 - \sin^2 \gamma_1 \sin^2 \theta_1}}, \quad \eta = \arccos \dfrac{\cos \gamma_2}{\sqrt{1 - \sin^2 \gamma_1 \sin^2 \theta_1}}. \quad \text{(S42)}$$

Based on the assignment of lattice points in the GFF sheet given in Fig. S6E, the lattice vectors corresponding to two different folding angle $\rho_1$ and $\rho_1{}'$ are

$$\mathbf{l}_1^{\rho_1} = \begin{pmatrix} L|_{\rho_1} \\ 0 \\ 0 \end{pmatrix}, \quad \mathbf{l}_2^{\rho_1} = \begin{pmatrix} 0 \\ W|_{\rho_1} \\ 0 \end{pmatrix}, \quad \mathbf{l}_3^{\rho_1} = \begin{pmatrix} -K|_{\rho_1} \\ -W|_{\rho_1}/2 \\ H|_{\rho_1} \end{pmatrix};$$

$$\mathbf{l}_1^{\rho_1'} = \begin{pmatrix} L|_{\theta_1'} \\ 0 \\ 0 \end{pmatrix}, \quad \mathbf{l}_2^{\rho_1'} = \begin{pmatrix} 0 \\ W|_{\rho_1'} \\ 0 \end{pmatrix}, \quad \mathbf{l}_3^{\rho_1'} = \begin{pmatrix} -K|_{\rho_1'} \\ -W|_{\rho_1'}/2 \\ H|_{\rho_1'} \end{pmatrix}. \quad \text{(S43)}$$

Since there is no diffusion, we can describe the lattice transformation due to folding from angle $\rho_1$ to $\rho_1{}'$ based on a transformation matrix $\mathbf{U}_{\text{3D-S}}$ such that

$$\left(\mathbf{l}_1^{\rho_1'} \quad \mathbf{l}_2^{\rho_1'} \quad \mathbf{l}_3^{\rho_1'}\right)^T = \mathbf{U}_{\text{3D-S}} \left(\mathbf{l}_1^{\rho_1} \quad \mathbf{l}_2^{\rho_1} \quad \mathbf{l}_3^{\rho_1}\right)^T. \quad \text{(S44)}$$

Here $\mathbf{U}_{\text{3D-S}}$ describes the lattice-distortive strains that transform the lattice; the subscript "3D-S" indicates three-dimensional shear. $\mathbf{U}_{\text{3D-S}}$ can be formulated as

$$\mathbf{U}_{\text{3D-S}} = \begin{pmatrix} L|_{\rho_{1'}} / L|_{\rho_1} & 0 & 0 \\ 0 & W|_{\rho_{1'}} / W|_{\rho_1} & 0 \\ \dfrac{K|_{\rho_1}\left(H|_{\rho_{1'}} / H|_{\rho_1}\right) - K|_{\rho_{1'}}}{L|_{\rho_1}} & \dfrac{W|_{\rho_1}\left(H|_{\rho_{1'}} / H|_{\rho_1}\right) - W|_{\rho_{1'}}}{2W|_{\rho_1}} & H|_{\rho_{1'}} / H|_{\rho_1} \end{pmatrix}. \tag{S45}$$

By substituting the expressions of $L$, $W$, $H$, and $K$ into Eq. (S45), one can correlate $\mathbf{U}_{\text{3D-S}}$ to the rigid folding kinematics.

$\mathbf{U}_{\text{3D-S}}$ is not a diagonal matrix. Its non-trivial main diagonal elements describes the contraction and (or) extension along the lattice vectors; and its non-zero off-diagonal elements describes the distortion due to shear. Note that the GFF origami is characterized by its ability to achieve out-of-plane shearing deformation during folding, which is characterized by the changes in angle $\psi$ (1). For example, by folding the GFF unit ($a = c$, $\gamma_1 = 36°$, $\gamma_2 = 72°$) from $\rho_1 = 90°$ to $\rho_1 = 30°$, the angle $\psi$ changes from $54.7°$ to $42.1°$. As a result, the out-of-plane shearing deformation is the kinematic origin of the 3D shear transformation.

**(8) 3D shuffle**

3D shuffle is observed when transforming a non-Bravais lattice of vertex inclusions based on stacked Miura-ori structure in a nested-in configuration ($0 \le \theta_A \le 90°$) shown in Fig. S6F. Construction of the stacked Miura-ori structure is introduced in Fig. S3B. Its folding can be described by the dihedral folding angle $\theta_A$ between a facet of the bottom Miura-ori unit and the reference *x-o-y* plane. In the nested-in configurations, the structure's outer dimensions are

$$H_A = c_A \sin\gamma_A \sin\theta_A, \quad H_B = c_B \sin\gamma_B \sin\theta_B, \quad H = H_B - H_A,$$
$$L = 2a_A \frac{\cos\theta_A \sin\gamma_A}{\sqrt{1 - \sin^2\gamma_A \sin^2\theta_A}}, \quad J = \frac{a_A}{\sqrt{1 + \tan^2\gamma_A \cos^2\theta_A}}, \quad W = 2c_A\sqrt{1 - \sin^2\gamma_A \sin^2\theta_A}. \tag{S46}$$

Unlike in the aforementioned three cases of 3D Bravais lattices, the three lattice vectors are not sufficient to describe the lattice here due to the additional lattice point inside the unit cell. Therefore, we need to incorporate the *shift* vector $\mathbf{p}$ in addition to the lattice vectors $\mathbf{l}_i$ (5), similar to the 2D shuffle study. The lattice vectors $\mathbf{l}_i$ describe the constituent Bravais lattice, and the shift vector $\mathbf{p}$ describes the offset between the two congruent lattices. Specifically, these vectors corresponding to two different folding angle $\theta_A$ and $\theta_A{}'$ can be formulated as

$$\mathbf{l}_1^\theta = \begin{pmatrix} L|_\theta \\ 0 \\ 0 \end{pmatrix}, \ \mathbf{l}_2^\theta = \begin{pmatrix} 0 \\ W|_\theta \\ 0 \end{pmatrix}, \ \mathbf{l}_3^\theta = \begin{pmatrix} 0 \\ 0 \\ H|_\theta \end{pmatrix}, \ \mathbf{p}^\theta = \begin{pmatrix} L|_\theta / 2 \\ J|_\theta + W|_\theta / 2 \\ H_A|_\theta \end{pmatrix};$$

$$\mathbf{l}_1^{\theta'} = \begin{pmatrix} L|_{\theta'} \\ 0 \\ 0 \end{pmatrix}, \ \mathbf{l}_2^{\theta'} = \begin{pmatrix} 0 \\ W|_{\theta'} \\ 0 \end{pmatrix}, \ \mathbf{l}_3^{\theta'} = \begin{pmatrix} 0 \\ 0 \\ H|_{\theta'} \end{pmatrix}, \ \mathbf{p}^{\theta'} = \begin{pmatrix} L|_{\theta'} / 2 \\ J|_{\theta'} + W|_{\theta'} / 2 \\ H_A|_{\theta'} \end{pmatrix}. \tag{S47}$$

Since there is no diffusion, we can then describe the lattice transformation due to folding from angle $\theta$ to $\theta'$ as a combination of deformation and shift, and derive the transformation matrix $\mathbf{T}_{\text{3D-SH}}$ such that

$$\left( \mathbf{l}_1^{\theta'} \quad \mathbf{l}_2^{\theta'} \quad \mathbf{l}_3^{\theta'} \quad \mathbf{p}^{\theta'} \right)^T = \mathbf{T}_{\text{3D-SH}} \left( \mathbf{l}_1^\theta \quad \mathbf{l}_2^\theta \quad \mathbf{l}_3^\theta \quad \mathbf{p}^\theta \right)^T. \tag{S48}$$

Here the subscript "3D-SH" means three-dimensional shuffle, and $\mathbf{T}_{\text{3D-SH}}$ can be written as

$$\mathbf{T}_{\text{2D-SH}} = \begin{pmatrix} & & & 0 \\ & \mathbf{U}_{\text{3D-D}} & & 0 \\ & & & 0 \\ \mu_1 & \mu_2 & \mu_3 & \lambda \end{pmatrix}. \tag{S49}$$

$\mathbf{T}_{\text{3D-SH}}$ is not a diagonal matrix, and it reflects how the lattice-distortive strain and shuffle component are integrated in the transformation. The submatrix $\mathbf{U}_{\text{3D-D}}$ relates to 3D dilation, which describes the lattice-distortive strains of the constituent 3D Bravias lattice. This submatrix can be replaced by other 3D transformation matrices (i.e., $\mathbf{U}_{\text{3D-S}}$, $\mathbf{U}_{\text{3D-C/E}}$, or $\mathbf{U}_{\text{3D-S}}$) if different origami structures are used. The non-trivial elements in the third row quantify the 3D shuffle component, and they satisfy the following relationship

$$\left( \mathbf{p}^{\theta'} \right)^T = \left( \mu_1 \quad \mu_2 \quad \mu_3 \quad \lambda \right) \left( \mathbf{l}_1^\theta \quad \mathbf{l}_2^\theta \quad \mathbf{l}_3^\theta \quad \mathbf{p}^\theta \right)^T. \tag{S50}$$

By substituting the expressions of $L$, $W$, $H$, $H_A$ and $J$ into Eq. (S49) and Eq. (S50), one can correlate $\mathbf{T}_{\text{3D-SH}}$ to the rigid-folding kinematics.

   Again, we emphasize that changing of the relative positions among vertices (or other characteristic entities) is ubiquitous during origami folding, and it is the fundamental mechanism that generates the shuffle during lattice transformation.

Overall, this section examines all types of 2D and 3D diffusionless lattice phase transformations by formulating the corresponding lattice transformation matrices based on the rigid-folding kinematics. Although the analyses here are based on specific examples of origami structures, the generic correlations between origami folding kinematics and lattice transformation are uncovered and summarized in Table S4.

# S4. Discrete symmetry switches

With carefully designed creases, the Miura-ori based 2D lattices of rods could reach four of the five 2D Bravais lattice types: namely rectangular (R), center-rectangular (CR), square (S), and hexagonal (H) lattices. The only exception is the oblique (O) lattice that requires a SC origami unit shown in Fig. S1A.

As mentioned in the main text, the correlations between origami geometries and lattice configurations are not unique. Especially, such correlations depend on the locations of assigned lattice points with respect to the different characteristic entities of origami, such as its vertices or crease lines. Here we consider a scenario that each vertex of the Miura-ori sheet is connected to a rod, which constitute a 2D lattice of rods. In what follows, we focus on how the 2D lattice of rods evolves with folding. To this end, two dimensionless parameter $A$ and $B$ are defined

$$ A = \frac{J}{L/2} = \frac{1}{\cos\theta\tan\gamma}, \quad B = \frac{W/2}{L/2} = \frac{c\left(1-\sin^2\theta\sin^2\gamma\right)}{a\cos\theta\sin\gamma} = \frac{c}{a}\cos\gamma\left(A+\frac{1}{A}\right). \tag{S51} $$

Hence, for each lattice type, the geometry relations among the lattice points can be expressed in terms of $A$ and $B$. Specifically, for any positive integer $n$, the rectangular lattice requires $A/B = n$; the center-rectangular lattice requires $A/B = n-1/2$; the square lattice requires $A = n$, $B = 1$; and the hexagonal lattice requires $A = (2n-1)/\sqrt{3}$, $B = 2/\sqrt{3}$ (6). Note that the hexagonal lattice is a special case of the center-rectangular lattice, and the square lattice is a special case of rectangular lattice.

Based on these relationships, different combinations of sector angle ($\gamma$) and crease length ratios ($c/a$) of the Miura-ori can be examined to determine whether they can achieve either rectangular lattice (Fig. S7A) or center-rectangular lattice (Fig. S7B) during rigid-folding. Note that the regions corresponding to rectangular lattice and center-rectangular lattice can overlap, indicating that the corresponding lattice is able to switch between these two types of lattices. Fig. 3A in the main text is a combination of Fig. S7A and Fig. S7B. Moreover, the regions corresponding to different values of $n$ can also overlaps, indicating the possibilities of multiple switches. When $\gamma$ and $c/a$ reach specific combinations, the center-rectangular lattice or rectangular lattice would become a hexagonal or square lattice, respectively (dashed curves in Fig. S7A and Fig. S7B). It is also possible to switch between hexagonal and square lattices, and between two different hexagonal lattices (bold curves H-S and H-S-H in Fig. 3A, main text). However, it is not possible to switch between two different square lattices.

In addition to the example in the main text, here we provide another example with $\gamma = 60°$ and $c/a = 0.7$ (shown as the square in the contour plot) to further illustrate the discrete symmetry switches. According to its location in the contour plots (Fig. S7A, B), this lattice can be switched among two rectangular lattices and two center-rectangular lattices by folding, and one of the center-rectangular lattices is also hexagonal. Simulations in Figure S7C illustrate these switches in details. The lattice switches to a rectangular lattice with $\mathbf{D_2}$ symmetry at folding angle $\theta = 39.2°$, then to a center-rectangular lattice with $\mathbf{D_2}$ symmetry at $\theta = 57.9°$, to another rectangular lattice with $\mathbf{D_2}$ symmetry at $\theta = 68.6°$, and finally to a hexagonal lattice with $\mathbf{D_6}$ symmetry at $\theta = 78.5°$.

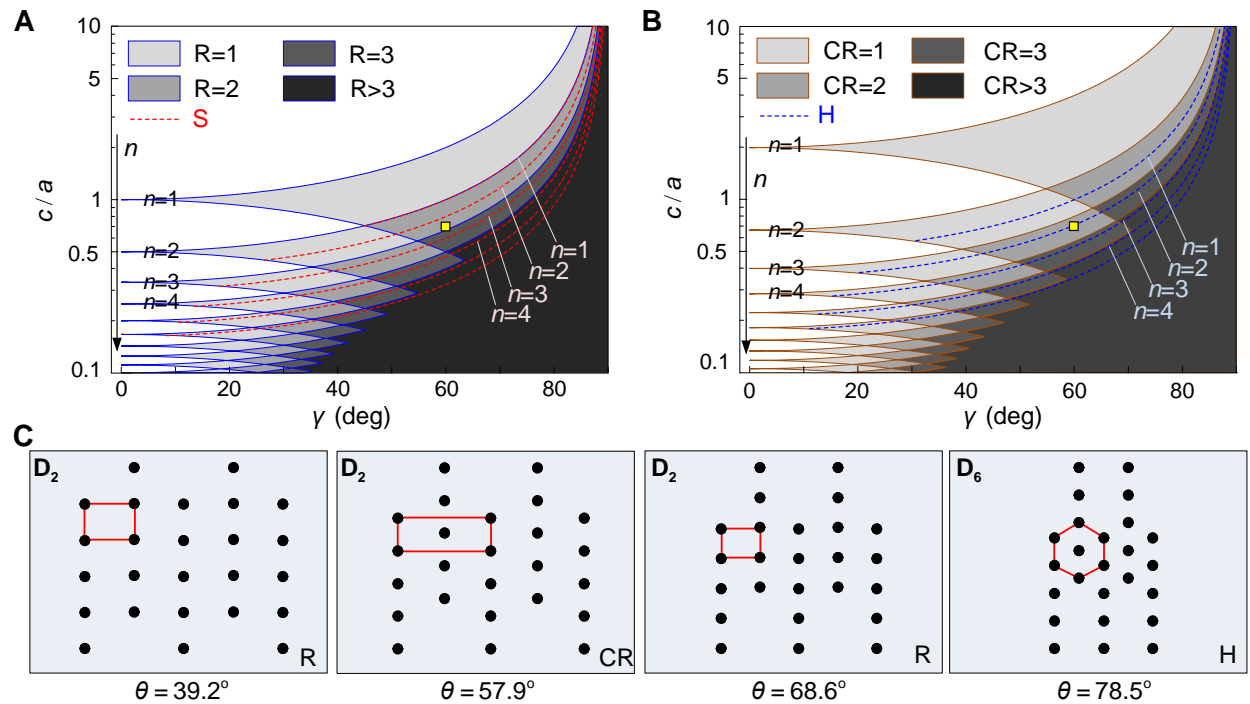

**Fig. S7. Discrete lattice switches in Miura-ori based 2D lattice of rods**. (**A**) The regions where rectangular lattices can be achieved; at the dashed lines, square lattices are possible; (**B**) The regions where center-rectangular lattices can be achieved; at the dashed lines, hexagonal lattices are possible; (**C**) Another example with $\gamma = 60°, c/a = 0.7$ (denoted by square in A and B). With this origami pattern, the lattice is able to switch from rectangular to center-rectangular, to rectangular, and to hexagonal.

## S5. Continuous symmetry measures and the Folding/Unfolding method

Here we provide a brief introduction to the definition and properties of the continuous symmetry measures (CSM) regarding origami lattices, as well as the procedures to evaluate the CSM values based on the *Folding/Unfolding* method. The underlying mathematical principles and the detailed deviations of the CSM can be found in (7, 8). We remark here the name "folding/unfolding" has no relation with the origami folding/unfolding, but indicate the symmetric operations applied on the points of the shape.

Figure S8A shows six distorted hexagons, and it is therefore interesting to evaluate how much symmetry each hexagon possesses with respect to a particular symmetry group. Here it is important to point out that we aim to quantify the distance of a given shape away from a symmetry group, and NOT the deviation from a specific predetermined shape with the desired symmetry. The CSM can (i) quantify the symmetry of a shape with respect to any symmetry group; (ii) determine the minimum CSM value this shape can exhibit and the corresponding symmetry group; and (iii) derive the corresponding closest shape that has the desired symmetry.

Given a shape consisting of $n_p$ points $P_i$ ($i = 1, ..., n_p$) and a symmetric group $\mathbf{G}$, the CSM $S(\mathbf{G})$ is a function of the minimum displacements by which $P_i$ have to undergo in order to acquire that $\mathbf{G}$ symmetry. We can use the Folding/Unfolding method (to be described below) to determine the points $\hat{P}_i$ ($i = 1, ..., n_p$) that constitute the closest shape with $\mathbf{G}$ symmetry, and then calculate the CSM value as follows

$$S(\mathbf{G}) = 100 \frac{1}{n} \sum_{i=1}^{n_p} \left\| P_i - \hat{P}_i \right\|^2. \tag{S52}$$

Here the squared values are used to ensure that the function is isotropic, continuous, and differentiable. Prior to evaluating the CSM value, the shape should be normalized by scaling about its centroid so that the maximum distance of any point $P_i$ to the centroid is 1. The factor of 100 ensures that the CSM value has a range of $0 \leq S(\mathbf{G}) \leq 100$. If a shape perfectly possesses the desired $\mathbf{G}$ symmetry, $S(\mathbf{G}) = 0$. $S(\mathbf{G})$ increases as the shape departs away from the $\mathbf{G}$ symmetry; and $S(\mathbf{G})$ reaches a maximum (not necessary 100) when the shape is farthest away from the $\mathbf{G}$ symmetry. The definition given above allows us to evaluate the CSM of any 2D or 3D shapes with respect to any symmetry group or symmetry element, and no reference shape is required *a priori*.

The purpose of Folding/Unfolding method is to construct a shape that is symmetric with respect to a given symmetry group. To detail the construction steps, we build a 2D shape with $\mathbf{D}_3$ symmetry as an example. The dihedral group $\mathbf{D}_3$ is made up of a rotation element $C_3$ and reflection element $\sigma$ (Fig. S8B), where the rotation is about the origin and the reflection is about one of the axes (say, $y$-axis). The $\mathbf{D}_3$ symmetry group is of order 6, that is, it contains 6 elements or symmetry operations that are described as follows (Fig. S8B):

$g_1 = E =$ the identity operation.
$g_2 = \sigma =$ reflection about the $y$-axis.
$g_3 = C_3 =$ rotation about the origin by $2\pi/3$ radians.
$g_4 = C_3\sigma = \sigma C_3^{\ 2} =$ rotation about the origin by $4\pi/3$ radians followed by reflection.
$g_5 = C_3^{\ 2} =$ rotation about the origin by $2\pi/3$ radians.
$g_6 = \sigma^2 C_3 = \sigma C_3 =$ rotation about the origin by $2\pi/3$ radians followed by reflection.

By applying the above six operations in sequence from $g_1$ to $g_6$, a 2D $\mathbf{D}_3$ symmetric shape can be generated based on an arbitrary point $P_1$ in the $x$-$o$-$y$ plane (Fig. S8C). Note that the operation sequence above is natural in that it considers the connectivity of the different points in the polygon. Many other operation sequences are possible to construct a $\mathbf{D}_3$ symmetric shape, but they do not necessarily guarantee the inherent connectivity of polygon points. Here, the procedures of generating a symmetric shape by applying a sequence of operations $g_i$ to a point are referred to as *unfolding*. The inverse procedures are called *folding*, i.e., points $P_2, P_3, \dots$ can be folded back to $P_1$ by applying the inversed operations $g_i^{-1}$ in sequence.

The Folding/Unfolding method is the basis for evaluating the CSM values of a shape with respect to a given symmetry group. Depending on whether the number of points ($n_p$) in the shape is equal to, larger than, or smaller than the number of elements in the symmetry group ($n_g$), four cases are possible: (I) $n_p = n_g$; (II) $n_p < n_g$, and $n_g = l n_p$, where $l = 2, 3, \dots$; (III) $n_p > n_g$, and $n_p = k n_g$, where $k = 2, 3, \dots$; (IV) $n_p \neq n_g$, and they cannot be expressed by $n_g = l n_p$ or $n_p = k n_g$ for any integers $l$ and $k$. In addition, we can have another case (V) in which there is a central point inside the polygon (for 3D, polyhedron). The method for evaluating CSM are different for these cases, but are all developed upon the basic case (I). Here, we focus on the basic case (I), in which the number of points ($n_p$) equals to number of elements in the symmetry group ($n_g$). As an example, we present the detailed procedures of evaluating the CSM of a distorted hexagon shown in Fig. S9 with respect to $\mathbf{D_3}$ symmetry. More details for other cases can be found in (8).

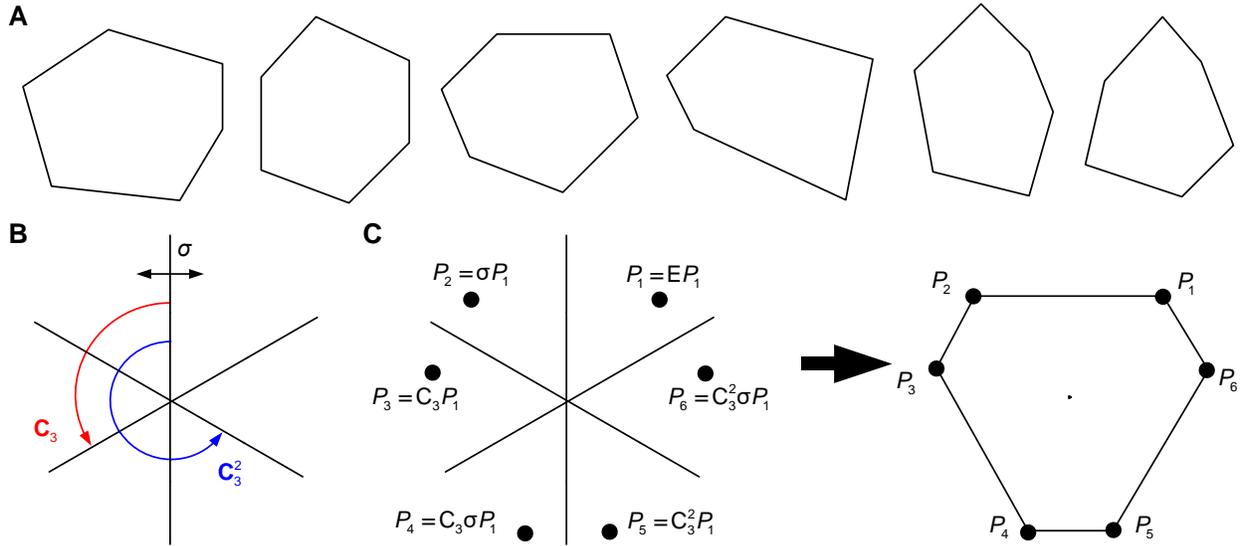

**Fig. S8. Creating a hexagon with $\mathbf{D_3}$ symmetry**. (**A**) Six distorted hexagon. (B) The $\mathbf{D_3}$ symmetry group is made up of the reflection operation ($\sigma$) and a rotation operation ($\mathbf{C_3}$), it contains six elements listed in the text. (**C**) Creating a $\mathbf{D_3}$-symmetric hexagon by applying a set of operations on a random point $P_1$ in specific sequence and connecting the points.

The CSM is evaluated as follows:

(1) Determine the centroid of the shape and translate it so that the centroid coincides with the origin of reference coordinates (Fig. S9A). Scale the shape so that the maximum distance from the origin to its points is 1 (Fig. S9B).

(2) List the elements of the targeted symmetry group, i.e., the symmetric operations $g_1, \ldots, g_{n_g}$. Translate these operations so that all rotations are about the origin, and all reflection lines pass through the origin. Select a sequence of the operations that follow the connectivity of the $n_p$ points of the shape. In this case, $n_p = n_g = 6$, and two possible sequences are $g_1 \rightarrow g_2 \rightarrow \ldots \rightarrow g_6$ and $g_6 \rightarrow g_5 \rightarrow \ldots \rightarrow g_1$.

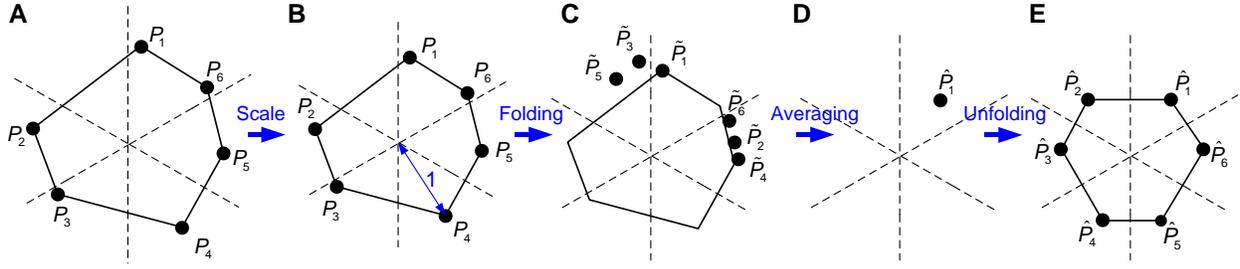

**Fig. S9. Evaluating the CSM of a distorted hexagon with respect to $\mathbf{D}_3$ symmetry based on the Folding/Unfolding method**. (**A**) Translate the centroid of the distorted hexagon to the origin. (B) Scale the hexagon so that the maximum distance from any point to the origin is 1. (**C**) Fold the points of the hexagon, and a cluster of points is obtained. (**D**) Average the cluster of points into an averaged point. (**E**) Unfold the averaged point into a shape with $\mathbf{D}_3$ symmetry.

(3) *Fold* the points $P_1, ..., P_{n_p}$ by applying the inverse symmetric operation $g_i^{-1}$ to the points $P_i$ in the sequence of $g_1^{-1} \to g_2^{-1} \to ... \to g_6^{-1}$, i.e., $\tilde{P}_i = g_i^{-1} P_i$ $(i=1,...,n_g)$. This gives rise to a cluster of points $\tilde{P}_1, ..., \tilde{P}_{n_g}$ (Fig. S9C).

(4) Average the folded points $\tilde{P}_i$ $(i=1,...,n_g)$ to obtain an averaged point $\hat{P}_1$ (Fig. S9D), i.e., $\hat{P}_1 = \dfrac{1}{n_g} \sum\limits_{i=1}^{n_g} g_i^{-1} P_i = \dfrac{1}{n_g} \sum\limits_{i=1}^{n_g} \tilde{P}_i$.

(5) *Unfold* the averaged point $\hat{P}_1$ by applying the $g_i$ operations in sequence, i.e., $\hat{P}_i = g_i \hat{P}_1$ $(i=1,...,n_g)$. This creates a set of new points $\hat{P}_1, ..., \hat{P}_{n_g}$ that constitute the reference shape with the desired $\mathbf{D}_3$ symmetry (Fig. S9E). Here the sequence of $g_i$ operations must be consistent with what is used in step (3) to retrieve the original connectivity of the shape.

(6) Calculate the CSM value according to Eq. (S52).

(7) Repeat steps (1) to (6) with all possible sequence (two sequences in this example) and identify the sequence that gives the minimum CSM value. The minimum CSM value corresponds to the best cluster of the folded points.

The minimum CSM value obtained through the procedures above is the minimum distance by which the points in this shape have to travel in order to acquire the desired symmetry. Proof of this conclusion can be found in (8).

The CSM plot in Fig. 4 of the main text is obtained by the Folding/Unfolding method. Specifically, for unit cell (ii), evaluation the CSM with respect to the $\mathbf{D}_2$ symmetry (i.e., $S_{ii}(\mathbf{D}_2)$ ) can be directly performed through the above listed steps, because the number of points in the parallelogram unit cell (ii) equals to the number of elements in the $\mathbf{D}_2$ symmetry group ( $n_p = n_g = 4$ ), so this belongs to the aforementioned case (I). However, when evaluating $S_{ii}(\mathbf{D}_4)$ and $S_i(\mathbf{D}_6)$, the number of elements in the symmetry group is larger than the number of points in the unit cell. For unit cell (ii) and symmetry group $\mathbf{D}_4$, $n_p = 4$, $n_g = 8$; for unit cell (i) and symmetry group $\mathbf{D}_6$, $n_p = 6$, $n_g = 12$. They both belong to case (II) since $n_g = l n_p$ $(l = 2)$. To deal with these scenarios, we treat each of the $n_p$ points as 2 repeated points. In this way, when calculating $S_{ii}(\mathbf{D}_4)$, we treat the $\mathbf{D}_4$-symmetric square as a degenerated octagon with each two points coinciding; when evaluating $S_i(\mathbf{D}_6)$, we treat the $\mathbf{D}_6$-symmetric hexagon as a degenerated dodecagon with each two points coinciding.